\documentclass[hidelinks,a4paper,10pt,twocolumn, switch]{article}

\pdfoutput=1

\usepackage{fancyhdr}
\fancypagestyle{plain}{%
  \fancyhf{}
\fancyfoot[R]{\thepage}
\fancyfoot[L]{\small \textit{Preprint submitted to Springer}\\ \texttt{arXiv}}
}

\usepackage[affil-it]{authblk}
\makeatletter
\def\@maketitle{%
  \newpage
  \null
  \vskip 2em%
  \begin{center}%
  \let \footnote \thanks
    {\Large\bfseries \@title \par}%
    \vskip 1.5em%
    {\normalsize
      \lineskip .5em%
      \begin{tabular}[t]{c}%
        \@author
      \end{tabular}\par}%
    \vskip 1em%
    {\normalsize \@date}%
  \end{center}%
  \par
  \vskip 1.5em}
\makeatother

\usepackage[a4paper, total={17.5cm,24cm}]{geometry}

\usepackage[font=normal,labelfont=bf]{caption}

\usepackage{sectsty} 

\usepackage{titlesec}
\titlespacing\section{0pt}{12pt plus 3pt minus 3pt}{1pt plus 1pt minus 1pt}
\titlespacing\subsection{0pt}{10pt plus 3pt minus 3pt}{1pt plus 1pt minus 1pt}
\titlespacing\subsubsection{0pt}{8pt plus 3pt minus 3pt}{1pt plus 1pt minus 1pt}

\titleformat{\section}{\normalfont\large\bfseries}{\thesection}{1em}{}

\titleformat{\subsection}{\normalfont\normalsize\bfseries}{\thesubsection}{1em}{}

\titleformat{\subsubsection}{\normalfont\normalsize}{\thesubsubsection}{1em}{}

\titleformat{\paragraph}[runin]{\normalfont\normalsize\itshape}{\theparagraph}{1em}{}

\usepackage[font={small}]{caption}

\usepackage{newtxtext,newtxmath} 

\usepackage[utf8]{inputenc}	
\usepackage[T1]{fontenc}	
\usepackage{xcolor}		
\usepackage{booktabs} 		
\usepackage{nicefrac}		
\usepackage{microtype}		
\usepackage{lineno}		
\usepackage{float}			

\usepackage{graphicx}
\usepackage{amsmath} 
\usepackage{newtxtext,newtxmath}      

\usepackage[hyperindex,breaklinks,linkcolor=blue,citecolor=blue,colorlinks=true]{hyperref}

\usepackage{enumitem} 
\usepackage{stmaryrd}  

\usepackage{algorithm}
\usepackage{algpseudocode}

\usepackage{siunitx}
\usepackage{breakcites}

\newcommand{\ind}[1]{{}^{#1}\!}
\newcommand{\dx}{\mathrm{d}}
\newcommand{\B}{\mathcal{B}}

\newcommand{\diffp}[2]{\frac{\partial #1}{\partial #2}}
\newcommand{\ve}[1]{\boldsymbol{#1}}
\newcommand{\te}[1]{\boldsymbol{#1}}
\newcommand{\bve}[1]{\bar{\boldsymbol{#1}}}
\newcommand{\bte}[1]{\bar{\boldsymbol{#1}}}

\newcommand{\V}[1]{\underline{\mathbf{#1}}}

\newcommand{\nablaX}{\nabla_{\!\!{\ve X}}}
\DeclareMathOperator{\tr}{tr}

\DeclareMathOperator{\Cof}{Cof}

\usepackage{mathtools}		
\mathtoolsset{centercolon}	

\newcommand{\Sym}{\mathscr{S\! y\! m}} 
\newcommand{\Orth}{\mathscr{O\! r\! t\! h}} 
\newcommand{\Softplus}{\mathscr{S\! P}}
 
\newcommand{\tte}[1]{{\mathbb{#1}}} 

\newcommand{\numbers}[1]{\mathbb{#1}}
\newcommand{\N}{\numbers{N}}    
\newcommand{\R}{\numbers{R}}    
\renewcommand{\L}[1]{\mathcal{L}_{#1}}

\usepackage{graphicx,psfrag,color}
\usepackage{tabularx}

\title{FE${}^\text{ANN}$ -- An efficient data-driven multiscale approach based on physics-constrained neural networks and automated data mining}

\begin{document}

\author{Karl A. Kalina, Lennart Linden, Jörg Brummund, Markus Kästner
  \thanks{Contact: \texttt{markus.kaestner@tu-dresden.de} }}
\affil{Institute of Solid Mechanics, \\ TU Dresden, Germany}

\date{}
\maketitle

\begin{abstract}
	Herein, we present a new data-driven multiscale framework called FE${}^\text{ANN}$ which is based on two main keystones: the usage of physics-constrained artificial neural networks (ANNs) as macroscopic surrogate models and an autonomous data mining process. Our approach allows the efficient simulation of materials with complex underlying microstructures which reveal an overall anisotropic and nonlinear behavior on the macroscale. Thereby, we restrict ourselves to finite strain hyperelasticity problems for now. 
	By using a set of problem specific invariants as the input of the ANN and the Helmholtz free energy density as the output, several physical principles, e.\,g., objectivity, material symmetry, compatibility with the balance of angular momentum and thermodynamic consistency are fulfilled a priori. 
	The necessary data for the training of the ANN-based surrogate model, i.\,e., macroscopic deformations and corresponding stresses, are collected via computational homogenization of representative volume elements (RVEs). 
	Thereby, the core feature of the approach is given by a completely autonomous mining of the required data set within an overall loop. 
	In each iteration of the loop, new data are generated by gathering the
	macroscopic deformation states from the macroscopic finite element (FE) simulation and a
	subsequently sorting by using the anisotropy class of the considered material. Finally, all unknown deformations are prescribed in the RVE simulation to get the corresponding stresses and thus to extend the data set. The proposed framework consequently allows to reduce the number of time-consuming microscale simulations to a minimum. It is exemplarily applied to several descriptive examples, where a fiber reinforced composite with a highly nonlinear Ogden-type behavior of the individual components is considered. Thereby, a rather high accuracy could be proved by a validation of the approach. 
	
\vspace{5mm}
\noindent
\textbf{Keywords: } Data-Driven Approach~\textendash~Artificial Neural Networks~\textendash~Anisotropic Hyperelasticity~\textendash~Computational Homogenization~\textendash~Decoupled Multiscale Scheme~\textendash~Physics-Constrained\\
\end{abstract}

\section{Introduction}
\label{sec:Intro}

Materials with an underlying meso- or microstructure, e.\,g., composites, solid foams, dual-phase steels or 3D-printed structures, enable the targeted design of engineering components with respect to their application. However, due to effects such as anisotropy, nonlinear or multiphysics phenomena, the experimental characterization of the effective constitutive behavior of such materials can be very complex.

\subsection{Multiscale schemes}

To avoid such a characterization of the material's effective behavior, computational \emph{multiscale schemes} can be used. These schemes allow the simulation of engineering components made of materials with underlying microstructure solely based on information about the microstructural arrangement and the properties of the individual components, e.\,g., matrix and inclusions. 
Basically, two different types of multiscale schemes exist: the \emph{coupled multiscale scheme} which is also known as FE${}^2$ approach (finite element square) \cite{Feyel2000,Miehe2002b,Schroder2014,Schroder2016,Keip2017,Lange2021,Koyanagi2021,Yvonnet2019} and the \emph{decoupled} or \emph{sequential multiscale scheme} \cite{Yamazaki2018,Terada2013,Kalina2020a,Yamamoto2022,Saito2021,Gebhart2022}. 

The FE${}^2$ method allows to completely couple the microscopic and macroscopic scales without the need for the explicit formulation of an effective constitutive model. It is thus universally applicable to arbitrary geometries if the necessary microscopic information are available. However, the decisive disadvantage is the very high computational effort, which results from the solution of the microscopic boundary value problem (BVP) for the homogenization at each integration point of the macroscopic FE mesh. 

Within a decoupled multiscale scheme, the material's effective behavior is initially determined from homogenization of representative volume elements (RVEs) and then a suitable constitutive model, the so-called macro or surrogate model, is calibrated by these data. With this model, macroscopic BVPs can now be solved, whereby the influence of the microstructure is implicitly captured by the macro model. Thus, in contrast to the FE${}^2$ method, the explicit solution of the microscopic BVP at each integration point is omitted.  The central disadvantage, however, is that the formulation of such a model can be extremely complicated.

\subsection{Data-based methods in solid mechanics}
To circumvent the time consuming task of formulating and calibrating a surrogate model within a decoupled multiscale scheme, \emph{data-based} or \emph{data-driven} techniques are very promising and have the potential to improve or replace traditional constitutive models. These techniques have become increasingly popular in the computational mechanics community during the last years \cite{Bock2019,Montans2019}.
In the following, a brief overview of the most common methods and their application to multiscale schemes is given.

\subsubsection{Overview on data-based constitutive modeling}
A relatively new strategy to substitute classical constitutive equations is the \emph{data-driven mechanics} approach, initially proposed by Kirchdoerfer~and~Ortiz~\cite{Kirchdoerfer2016} and extended to, e.\,g., noisy data sets \cite{Kirchdoerfer2017}, finite strains \cite{Nguyen2018}, inelasticity \cite{Eggersmann2019} or fracture mechanics \cite{Carrara2020} in the meantime. This method completely avoids to use constitutive equations. Instead, sets of stress-strain tuples which characterize the material’s behavior are used. A data-driven solver hence seeks to minimize the distance between the searched solution and the material data set within a proper energy norm, while compatibility and equilibrium have to be satisfied simultaneously.

The construction of so called \emph{constitutive manifolds} from collected data is an alternative approach which is described for elasticity and inelasticity by Iba\~{n}ez et al.~\cite{Ibanez2018}. With this technique, it is possible to strictly fulfill the 2nd law of thermodynamics, i.\,e., the thermodynamic consistency,  by using the \emph{GENERIC} paradigm (General Equation for Non-Equilibrium Reversible-Irreversible Coupling) during the construction of constitutive manifolds \cite{Gonzalez2019}.

Besides the previously mentioned techniques, there exist numerous data-based methods originating from the field of \emph{machine learning (ML)}. Probably, the most common technique is the application of \emph{artificial neural networks (ANNs)}, which have already been proposed in the early 90s by the pioneering work of Ghabussi~et~al.~\cite{Ghaboussi1991}.
In the last decades, ANNs have been intensively used for mechanical material modeling and simulations by means of the finite element method (FEM), e.\,g., in \cite{Hashash2004,Zopf2017,Stoffel2018,Tac2021,Kalina2021} among others. 

However, in general, a large amount of data is required to train ANNs to serve as robust and accurate surrogate models for systems with complex underlying physics, e.\,g., constitutive models. In this context, a comparatively new branch of ML techniques related to constitutive modeling are approaches classified as \emph{physics-informed, physics-constrained} \cite{Frankel2020,Fuhg2022}, \emph{mechanics-informed} \cite{Asad2022} or \emph{hybrid models} \cite{Settgast2020,Malik2021}.\footnote{In the following, the term physics-informed does not directly refer to the PINN-approach (physics-informed neural network) according to Raissi~et~al.~\cite{Raissi2019}.
	In a PINN, the searched solution field approximated by the ANN, e.\,g., displacement $\ve u(\ve x,t)$, is inserted into the governing partial differential equation (PDE) at collocation points. This expression is then added to the loss, so that the fulfillment of the PDE is enforced. In the context of constitutive modeling, however, the idea to enrich the ML approach with physical knowledge is applied in a similar way.}
In these methods, essential physical principles and information are inserted into the ML-based model or parts of classical models are replaced with data-based methods, which leads to an improvement of the extrapolation capability and enables training with sparse data. 
In the case of ANNs, this can be achieved via the network architecture or by adapted training algorithms. By choosing problem-specific invariant sets as the input variables, material symmetries are automatically satisfied for hyperelastic models \cite{Linka2021,Kalina2021,Klein2021}. Furthermore, the \emph{thermodynamic consistency} can be fulfilled by choosing the free energy as the output quantity. In order to train the ANN with respect to the stresses, gradients of the output with respect to the input are inserted into the loss  \cite{Fernandez2020a,Linka2021}. This technique is also named as Sobolev training in \cite{Vlassis2020}.
Furthermore, physical knowledge can be inserted via constraint training processes \cite{Weber2021}. 
For the consideration of dissipative behavior, Masi et al.~\cite{Masi2021} proposed an adapted ANN architecture consisting of two feed forward neural networks (FNNs).
Thereby, the first network is used to predict the evolution of internal variables and the second for approximating the free energy. 
Finally, the combination of classical models with data-based techniques is a further method to insert physical knowledge. Thereby, only parts of a model are replaced by data-based techniques, e.\,g., in plasticity models \cite{Settgast2020,Malik2021,Vlassis2021}.

\subsubsection{Data-based multiscale modeling and simulation}
In the context of \emph{multiscale problems}, the mentioned data-based methods can be used as \emph{surrogate models} which replace the computationally expensive simulation of RVEs \cite{Liu2021}. The high flexibility of trained networks with simultaneously excellent prediction qualities is thereby proven in numerous works:
In \cite{Fernandez2020a} ANNs are used to describe the homogenized response of cubic lattice metamaterials exhibiting large deformations and buckling phenomena. Thereby, several types of hyperelastic ML-based models which fulfill basic physical requirements are used and compared to each other. Based on this, an extension to polyconvex hyperelastic models is given in \cite{Klein2021}. 
In \cite{Li2020a}, \emph{FFNs} and \emph{RNNs (recurrent neural networks)} are used to replicate the homogenization of RVEs revealing inelastic behavior of the individual components. Thereby, also unknown paths can be predicted with the trained networks. Similarly, ANNs that replace  the inelastic constitutive responses of composite materials sampled by RVE simulations are shown in \cite{Fuchs2021}. Thereby, in addition, a deep reinforcement learning combinatorics game is used to automatically find an optimal set of network hyper-parameters from a decision tree. 
A data-driven multiscale framework called \emph{deep material network} is shown in \cite{Liu2019a,Liu2019,Gajek2020}. Thereby, the homogenized RVE response is reproduced by a network including a collection of connected mechanistic building blocks with analytical homogenization solutions. With that, a complex effective response could be described without the loss of essential physics. An extension of the deep material network approach to fully coupled thermo-mechanical multiscale simulations of composite materials is given in \cite{Gajek2022}.

Furthermore, ANNs can be used to link \emph{microstructural characteristics} and \emph{effective properties}.
For example, in \cite{Fuhg2022a}, an ML framework for predicting macroscopic yield as a function of crystallographic texture is described. An ML-based multiscale calibration of constitutive models representing the effective response of rate dependent composite materials is applied to brain white matter in \cite{Field2021}. Based on a library of calibrated parameters corresponding to a set of microstructural characteristics, an ML model which predicts the constitutive model parameters directly from a new microstructure is trained. Similarly, an ANN is trained to predict the elastic properties of short fiber reinforced plastics in \cite{Breuer2021}. 

Finally, the application of \emph{data-based} techniques as \emph{surrogate models} in \emph{decoupled multiscale} \emph{schemes}, which enable the simulation of macroscopic engineering components, is promising.	
In the pioneering work \cite{Le2015}, a decoupled multiscale scheme using an ANN-based surrogate model has been shown for elastic composites with cubic microstructures.
Another multiscale methodology is presented in \cite{Fuhg2021}. Therein, a hybrid macroscopic surrogate model is used, i.\,e., a traditional constitutive model is combined with a data-driven correction.
An ML-based multiscale framework for the simulation of the elastic response of metals having a polycrystalline microstructure is presented in \cite{Vlassis2020}. Therein, the database is generated by using a 3D \emph{FFT (fast Fourier transform)} solver. Based on this, \cite{Vlassis2021b,Vlassis2021} show an extension to elastoplasticity, whereby ANNs are used for the description of the yield surface and the stress within a hybrid modeling approach. Similarly, the simulation of the elastic-plastic deformation behavior of open-cell foam structures is shown in \cite{Settgast2020,Malik2021}. Thereby, a hybrid model including two ANNs is used as the macroscopic surrogate model. Therein, the first network serves for the description of the macroscale yield function and the second one for the prediction of the floating direction. In \cite{Gajek2022}, two scale simulations of thermo-mechanical problems are solved by using deep material networks.	
Applying the modeling strategy initially proposed in \cite{Masi2021}, a multiscale scheme for the inelastic behavior of lattice material structures is described in \cite{Masi2021a}. An application of ANNs as surrogate models describing the  anisotropic electrical response of graphene/polymer nanocomposites is shown in \cite{Lu2019}. Furthermore, in \cite{Wu2020}, the application of recurrent neural networks (RNNs) as surrogate models within multiscale simulations of elastoplastic problems is shown. Furthermore, the RNN-based approach is compared to full FE${}^2$ simulations. An application of RNN surrogate models to viscoplasticity is described in \cite{Ghavamian2019}. A combination of fully coupled FE${}^2$ simulations with an adaptive switching to ANN-based surrogate models for the complex simulation of RVEs is shown in \cite{Fritzen2019}.

In addition, a multiscale framework based on the \emph{data-driven mechanics} approach \cite{Kirchdoerfer2016} is presented in \cite{Karapiperis2021} for the application to sand.
Thereby, the necessary data set is extracted from lower scale simulations. In \cite{Korzeniowski2021}, a multi-level method is used within a data-driven multiscale scheme which is applied for the simulation of solid foam materials.

\subsection{Content}
Within this contribution, an efficient data-driven multiscale approach called \emph{FE}${}^\textit{ANN}$, which makes use of \emph{phy\-sics-con\-strai\-ned ANNs} as a macroscopic surrogate model, is presented. The approach allows to consider materials with complex microstructures leading to an overall anisotropic behavior, whereby a restriction to hyperelastic solids is made for now. The ANN-based surrogate model automatically fulfills several physical principles, e.\,g., \emph{objectivity, material symmetry}, compatibility with the \emph{balance of angular momentum} and the \emph{thermodynamic consistency}. This is done by using a set of problem specific invariants $I_k$ as the input of the network and the Helmholtz free energy density $\psi$ as the output, cf. Linka~et~al.~\cite{Linka2021}, Klein~et~al.~\cite{Klein2021} or Linden~et~al.~\cite{Linden2021}, among others. 
The data which are used to train the ANN are collected via computational homogenization of an RVE representing the material's microstructure.	Thereby, in contrast to most of the data-driven multiscale approaches from the literature, it is not necessary to explore all required macroscopic states of deformation which occur in the macroscopic simulation in advance. Instead, the required data, i.\,e., effective deformations and corresponding stresses, are determined by homogenization in a fully \emph{autonomous} way within the framework by collecting the relevant deformations from the macroscopic FE simulation. This procedure is similar to the approach presented by Korzeniowski~and~Weinberg~\cite{Korzeniowski2021} which is based on the data-driven mechanics approach \cite{Kirchdoerfer2016}.
Here, in addition, the macroscopic deformations are mapped into an invariant space associated to the symmetry group of the considered material. In this space, the selection of relevant states is done so that the number of time-consuming microscale simulations can be reduced to a minimum. Moreover, it is not necessary to perform a rough scan of the relevant deformation area in advance here.
The proposed framework is exemplarily applied to several descriptive examples, where a fiber reinforced composite with a highly nonlinear behavior of the individual components is considered. 

The organization of the paper is as follows: In Sect. \ref{sec:2}, the basic equations of the finite strain continuum solid mechanics theory as well as basic principles of hyperelastic models are given.  
After this, the proposed data-driven multiscale framework is described in Sect. \ref{sec:3}. This approach is exemplarily applied within several numerical examples in Sect. \ref{sec:4}. After a discussion of the results, the paper is closed by concluding remarks	and an outlook to necessary future work in Sect. \ref{sec:Conc}.

\section{Continuum solid mechanics}
\label{sec:2}
In this section, the basic kinematic and stress quantities as well as general relations of anisotropic hyperelastic constitutive models are summarized shortly. The reader is referred to the  textbooks of Haupt~\cite{Haupt2000}, Holzapfel~\cite{Holzapfel2000} or Ogden~\cite{Ogden1997} for a detailed overview. Furthermore, a Hill-type homogenization framework is introduced. For a clear mathematical notation, the space of tensors 
\begin{align}
	\L{n}:=\underbrace{\R^3 \otimes \cdots \otimes \R^3}_{n\text{-times}} \ \forall n\in\N_{\ge 1} \; , \label{eq:tensorSpace}
\end{align}
except for a tensor of rank zero, is used in the following. In Eq.~\eqref{eq:tensorSpace}, $\R^3$, $\N$ and $\otimes$ denote the Euclidean vector space, the set of natural numbers and the dyadic product, respectively. Tensors of rank one and two are given by boldface italic symbols in the following, i.\,e., $\ve A \in \L{1}$ or $\te B,\te C \in \L{2}$. Furthermore, a single or double contraction of two tensors is given by $\te B \cdot \te C = B_{kq} C_{ql} \ve e_k \otimes \ve e_l$ or $\te B:\te C=B_{kl}C_{lk}$, respectively. Thereby, $\ve e_k\in \L{1}$ denotes a Cartesian basis vector and the Einstein summation convention is used.

\subsection{Kinematics and stress measures}
\subsubsection{Kinematics}
In the following, a material body $\mathcal C$ which occupies the reference configuration \mbox{$\B_0 \subset \R^3$} at time $t_0 \in \R_{\geq 0}$ and the current configuration \mbox{$\B \subset \R^3$} at time \mbox{$t\in \mathcal T:=\{\tau\in\R_+ \,|\,\tau \ge t_0\}$} is considered. The displacement vector $\ve u \in \L{1}$ of a material point $P\in\mathcal C$ capturing the positions $\ve X \in \B_0$ at $t_0$ and $\ve x \in \B$ at $t$ is given by \mbox{$\ve u(\ve X,t):=\ve \varphi(\ve X,t) - \ve X$}. Therein, $\ve \varphi: \B_0 \times \mathcal T \to \B\; , (\ve X, t) \mapsto \ve x=:\ve \varphi(\ve X,t)$ denotes a bijective motion function which is postulated to be continuous in space and time. As further kinematic quantities, the deformation gradient $\te F \in \L{2}$ and the Jacobian determinant  $J\in\R_+$ are defined by the relations
\begin{align}
	\te F := (\nablaX \ve \varphi)^\text{T} \quad \text{and} \quad
	J:=\det \te F > 0 \; .
\end{align}
In the equation above, $\nablaX$ is the nabla-operator with respect to reference configuration $\B_0$. 

A deformation measure which is free of rigid body motions is given by the positive definite right Cauchy-Green deformation tensor \mbox{$\te C:=\te F^\text{T} \cdot \te F \in \Sym$} with the space of symmetric second order tensors \mbox{$\Sym:=\left\{\te \tau \in \L{2} \, |\, \te \tau = \te \tau^\text{T}\right\}$}. With that and by using Sylvesters formula, the spectral decomposition of $\te C$ follows to
\begin{align}
	\label{eq:projection}
	\te C = \sum_{\beta =1}^{N_\lambda} \lambda_\beta^2 \te P^\beta \; \text{with} \; 
	\te P^\beta := \delta_{1 N_\lambda} \te 1
	+ \prod_{\beta \ne \alpha}^{N_\lambda} \frac{\te C - \lambda_\beta^2 \te 1}{\lambda_\alpha^2 - \lambda_\beta^2}
	\; ,
\end{align}
where $\te 1\in\L{2}$, $\lambda_\beta\in\R_{\geq 0}$ and $\te P^\beta \in \Sym$ denote the identity tensor, the principal stretches and the projection tensors, respectively. The introduced symbol \mbox{$N_\lambda\in\{1,2,3\}$} indicates the algebraic multiplicity of $\lambda_\beta$. Furthermore, $\delta_{kl}$ is defined as the Kronecker delta. 

\subsubsection{Stress measures}
Within the framework of nonlinear continuum solid mechanics, various stress measures can be defined. The symmetric Cauchy stress $\te \sigma \in \Sym$, also known as true stress, is defined with respect to the current configuration $\B$. 
Furthermore, the 1st and 2nd Piola-Kirchhoff stress tensors $\te P \in \mathcal L_2$ and $\te T \in \Sym$ follow from the pull back operations
\begin{align}
	\te P := J \te F^{-1}\cdot\te \sigma \quad \text{and} \quad
	\te T := J \te F^{-1}\cdot\te \sigma \cdot \te F^{-\text{T}} \; .
\end{align}	
Accordingly, $\te P$ is related to both, $\mathcal B$ and $\mathcal B_0$, whereas $\te T$ is completely defined with respect to $\mathcal B_0$.

\subsection{Hyperelasticity}
\label{subsec:Hyper}
\subsubsection{General properties}
The constitutive behavior of the considered solids is restricted to \emph{hyperelasticity} within this work. Accordingly, a hyperelastic potential which is equal to the  Helmholtz free energy density function \mbox{$\psi: \L{2} \to \R_+\; , \te F \mapsto \psi(\te F)$} exists. By applying the procedure of Coleman~and~Noll~\cite{Coleman1963}, the relations 
\begin{align}
	\label{eq:consistent}
	\te P = \diffp{\psi}{\te F^\text{T}} \; \text{and} \;  \te T = \diffp{\psi}{\te F^\text{T}} \cdot \te F^{-\text{T}} \; .
\end{align}
then follow from the evaluation of the Clausius-Planck inequality \cite{Holzapfel2000}. 
With that, the \emph{thermodynamic consistency} of any hyperelastic model is fulfilled a priori.

In addition, there are some further requirements on $\psi$, which ensure a physically reasonable constitutive behavior \cite{Holzapfel2000,Ogden1997}.
The \emph{normalization condition }requires that  \mbox{$\psi(\te 1)=0$}, i.\,e., that the free energy vanishes in the undeformed state. Furthermore, the free energy should increase in any case if deformation appears: \mbox{$\psi(\te F)>0 \, \forall \te F\ne\te 1$}. The former two conditions imply that $\psi$ has a global minimum at \mbox{$\te F = \te 1$}. Consequently, the undeformed state is stress-free, i.\,e., \mbox{$\te T(\te 1) = \te 0$} holds. Additionally, the \emph{growth condition} requires that an infinite amount of energy is needed to infinitely expand the volume or compress it to zero: \mbox{$\psi(\te F)\to\infty$ as $(J\to\infty \vee J \to 0^+)$}. Finally, the principle of \emph{material objectivity} states that the free energy is invariant with respect to a superimposed rigid body motion. This statement is expressed by the relation \mbox{$\psi(\te F) = \psi(\te Q \cdot \te F)$} which holds for all special orthogonal tensors
\mbox{$\te Q \in \Orth^+:=\{\te \tau \in \L{2} \, | \, \te \tau^\text{T} = \te \tau^{-1}, \, \det \te \tau \equiv 1\}$}. Accor\-ding\-ly, the principle of material objectivity is automatically fulfilled if the tensor $\te C$ is used as the argument of $\psi$ instead of $\te F$, i.\,e., \mbox{$\psi: \Sym \to \R_+\; , \te C \mapsto \psi(\te C)$}, which is done in the following. 

Finally, \emph{polyconvexity} of the energy $\psi$, i.\,e., convexity with respect to $\te F$, $\Cof \te F := J \te F^{-\text{T}}$ and $\det \te F$, is a further condition. This condition implies material stability, i.\,e., Legendre-Hadamard-Ellipticity, see Ebbing~\cite{Ebbing2010} for more details. However, polyconvexity is a quite strong requirement on the free energy.  

\subsubsection{Material symmetry}
\label{subsubsec:MaterialSymmetry}

Besides the previously mentioned requirements, the constitutive equations should also reflect the symmetry of the described material which is expressed by the \emph{principle of material symmetry} \cite{Haupt2000,Ebbing2010}. According to that, the free energy have to be invariant with respect to all orthogonal transformations belonging to the symmetry group $\mathscr G$ of the underlying material: $\psi(\te C) = \psi(\te Q \cdot \te C \cdot \te Q^\text{T})$ for all $\te Q \in \mathscr G \subset \Orth$ with \mbox{$\Orth:=\{\te \tau \in \L{2} \, | \, \te \tau^\text{T} = \te \tau^{-1}, \, \det \te \tau \equiv \pm 1\}$}.

In order to describe anisotropic constitutive behavior, the concept of structural tensors can be used \cite{Haupt2000,Ebbing2010}. Depending on the considered material, these tensors are of order two \mbox{$\te M^1,\te M^2,\dots,\te M^{n_2} \in \L{2}$}, four \mbox{$\tte M^1,\tte M^2,\dots,\tte M^{n_4} \in \L{4}$}, or even higher. They reflect the material's anisotropy and are thus invariant with respect to the symmetry transformations, e.\,g.,
\begin{align}
	\te M^\alpha = \te Q \cdot \te M^\alpha \cdot \te Q^\text{T} \; \text{and} \; \tte M^\beta = \te Q * \tte M^\beta \, \forall \te Q\in\mathscr G \; .
\end{align}
The notation \mbox{$\te Q * \tte M^\beta$} means \mbox{$Q_{IM}Q_{JN}Q_{KO}Q_{LP} M_{MNOP}^\beta$}, where the Einstein summation convention is used. If the structural tensors are appended to the list of arguments of $\psi$, the energy is an isotropic tensor function even if the material is anisotropic which means that
\begin{align}
	\psi(\te C, \mathcal M_2, \mathcal M_4) = \psi(\te Q \cdot \te C \cdot \te Q^\text{T}, \te Q \cdot \mathcal M_2 \cdot \te Q^\text{T},\te Q *\mathcal M_4)
	\label{eq:isotropicFunctionA}
\end{align}
holds for all $\te Q \in \Orth$. To abbreviate the notation, the sets \mbox{$\mathcal M_2 :=\{\te M^1,\te M^2,\dots,\te M^{n_2}\}$} and \mbox{$\mathcal M_4 :=\{\tte M^1,\tte M^2,\dots,\tte M^{n_4}\}$} have been used in Eq.~\eqref{eq:isotropicFunctionA}.

Finally, a set \mbox{$\V I :=(I_1,I_2,\dots,I_n) \in \R^{n\times 1}$} consisting of \mbox{$n\in \N$} irreducible scalar valued invariants $I_\alpha \in \R$ could be derived by using the Cayley-Hamilton theorem.\footnote{The Cayley-Hamilton theorem states that a second order tensor fulfills his own eigenvalue equation, e.\,g., \mbox{$\te C^3 - I_1 \te C^2 + I_2 \te C -I_3 \te 1 = \te 0$}, where $I_1,I_2,I_3$ denote the principal invariants of $\te C$, cf. Eq.~\eqref{eq:isotropicInvariants}. Consequently, any power $\te C^n$ with $n\ge 3$ as well as the inverse $\te C^{-1}$ are expressible in terms of $\te C^2$, $\te C$ and $\te 1$ \cite{Haupt2000}.} Consequently, the free energy is expressed by the isotropic tensor function $\psi=\psi(I_1,I_2,\dots,I_n)$ which is invariant with respect to all transformations $\te Q \in \Orth$. By using the latter definition and applying the chain rule, the 2nd Piola-Kirchhoff stress $\te T$ follows to 
\begin{align}
	\te T = \sum_{\alpha=1}^{n}2 \diffp{\psi}{I_\alpha} \underbrace{\diffp{I_\alpha}{\te C}}_{\displaystyle =:\te G^\alpha} \; , \label{eq:Comp}
\end{align}
where $\te G^\alpha \in \Sym$ denotes tensor generators corresponding to the invariants $I_\alpha$ \cite{Kalina2021}.

\subsubsection{Special anisotropy classes}
\label{SubSubSec:AnisotropyClass}

Within this work, two anisotropy classes are considered, isotropic as well as transversely isotropic materials. Corresponding sets of irreducible invariants are given in the following. 

A set of irreducible invariants describing an \emph{isotropic hyperelastic} solid is given by \mbox{$\V I^\circ := (I_1,I_2,I_3) \in \R^{3\times 1}$}. The used principal invariants are given by
\begin{align}
	I_1:=\det \te C,\; I_2:=\tr (\Cof \te C) ,\;I_3:=\det \te C \; ,
	\label{eq:isotropicInvariants}
\end{align}
whereby the cofactor of $\te C$ is defined by \mbox{$\Cof \te C := J^2 \te C^{-1}$}.\footnote{Note that the more common expression $I_2 = \frac{1}{2} (\tr^2 \te C - \tr \te C^2)$ is equivalent to $I_2=\tr (\Cof \te C)$. Both expressions can be transformed into each other by using the Cayley-Hamilton theorem.}
Note that $\V I^\circ$ is also expressible by the principal stretches $\lambda_\alpha$ which follows from Eq.~\eqref{eq:projection}:
\begin{align}
	I_1 = \sum_{\alpha=1}^{N_\lambda} \nu_\alpha \lambda_\alpha^2 \, , \,
	I_2 = \prod_{\alpha=1}^{N_\lambda} \lambda_\alpha^{2 \nu_\alpha} 
	\sum_{\beta=1}^{N_\lambda} \nu_\beta \frac{1}{\lambda_\beta^{2}}\, \text{,} \, 
	I_3 = \prod_{\alpha=1}^{N_\lambda} \lambda_\alpha^{2 \nu_\alpha} \, .
	\label{eq:InvariantsStretches}
\end{align}

According to \cite{Haupt2000,Holzapfel2000,Ebbing2010}, the structural tensor describing \emph{transverse isotropy} is given by the second order tensor \mbox{$\te M = \ve A \otimes \ve A$}, whereby \mbox{$\ve A \in \L{1}$ with $|\ve A| \equiv 1$} is the fiber direction in the undeformed state. A set of irreducible invariants is thus \mbox{$\V I^\shortparallel:=(I_1,I_2,I_3,I_4,I_5)\in \R^{5\times 1}$}. Therein, the latter two invariants, which capture the material's anisotropy, are given by the expressions
\begin{align}
	I_4 := \te M : \te C \; \text{and} \; I_5 := \te M : \te C^2 \; .
	\label{eq:transverseIsotropicInvariants}
\end{align}

\subsection{Scale transition scheme}
\label{subsec:Homogenization}

\begin{figure}
	\centering
	\includegraphics{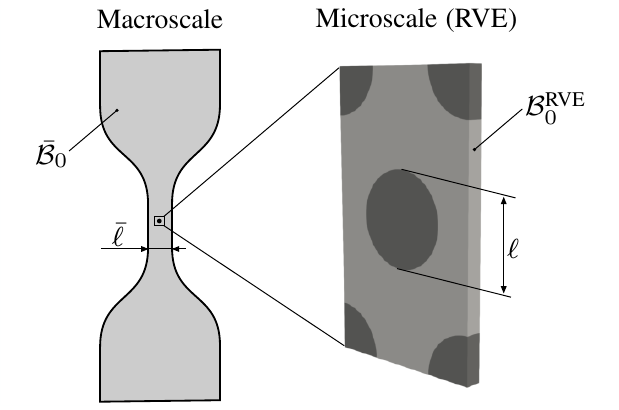}
	\caption{Schematic depiction of micro- and macroscale within a multiscale problem. To enable scale separation $\bar \ell \gg \ell$ must hold.}
	\label{fig:1}
\end{figure}

In the following, a distinction between two different scales, the \emph{micro-} and the \emph{macroscale} is made. The former is characterized by a heterogeneous structure consisting of matrix $\B_0^\text{m}\subset \R^3$ and inhomogeneities $\B_0^\text{i}\subset \R^3$ of characteristic length $\ell\in \R_+$, whereas the second considers a macroscopic body $\bar \B_0 \subset \R^3$ of characteristic length $\bar \ell \in \R_+$ and is assumed to be homogeneous. For the lengths introduced, the relation $\bar{\ell} \gg \ell$ known as scale separation must hold \cite{Schroder2014}. To label macroscopic quantities, they are marked by $\bar{(\bullet)}$ in the following.

In order to connect microscopic and macroscopic tensor quantities, an appropriate homogenization scheme is needed. Consequently, each macroscopic point $\bve{X} \in \bar\B_0$ gets assigned properties resulting from the behavior of the microstructure.	For this purpose, a \emph{representative volume element (RVE)} of the material is considered on the microscale in the vicinity of $\bve X$, cf. Fig.~\ref{fig:1}. An effective macroscopic quantity is then identified from the microscopic field distribution within the RVE by the volume average
\begin{align}
	\langle (\bullet) \rangle := \frac{1}{V^\text{RVE}}\int\limits_{\B_0^\text{RVE}} (\bullet) \, \dx V \; .
	\label{eq:average}
\end{align}
Using Eq. \eqref{eq:average}, the macroscopic deformation gradient and the 1st Piola-Kirchhoff stress are defined by	$\bte F := \langle \te F \rangle$ and $\bte P := \langle \te P \rangle$, respectively \cite{Schroder2014,Schroder2016,Kalina2020a}. 
Appropriate boundary conditions for the microscopic BVP, which has to be solved before the volume averaging can be performed, are deducible from the equivalence of the macroscopic and the averaged microscopic energies which is also known as the Hill-Mandel condition. For the considered finite strain setting it is given by the following relation \cite{Schroder2014,Schroder2016,Kalina2020a}: 
\begin{align}
	\bte P : \dot{\bte F} = \langle \te P : \dot{\te F} \rangle \; , \label{eq:HillMandel}
\end{align}
where $\dot{(\bullet)}$ denotes the material time derivative. Regarding purely hyperelastic behavior, the \emph{Hill-Mandel condition} expresses the equality of the rates of the macroscopic and the averaged microscopic Helmholtz free energies: $\dot{\bar \psi} = \langle \dot \psi \rangle$. Consequently, it holds $\bar \psi = \langle \psi \rangle$.

To fulfill Eq. \eqref{eq:HillMandel}, several type of boundary conditions (BCs) can be used, whereby \emph{periodic BCs} given by the spaces
\begin{align}
	\ve u\in
	\mathcal U(\bar{\te F}) &:= \left\{ \ve u \in \R^3 \; | \;
	\ve u = (\bar{\te F} - \te 1) \cdot \ve X + \tilde{\ve u}, \; \tilde{\ve u}^+ = \tilde{\ve u}^-\right\} 
	\; \text{,}
	\label{eq:periu1}\\ 
	\ve p \in \mathcal P &:= \left\{ \ve p \in \R^3 \; |\; \ve p^- = -\ve p^+ \right\}
	\label{eq:periu2}
\end{align}
are applied within this work \cite{Kalina2020a}. In the equation above, \mbox{$\ve p = \ve N \cdot \te P$} is the nominal stress vector. Furthermore, $(\bullet)^+$ and $(\bullet)^-$ denote values on opposing boundaries of the RVE which is supposed to be periodic. The tilde $\tilde{(\bullet)}$ marks the fluctuation part of a microscopic tensor quantity. 

\section{ANN-based multiscale approach with autonomous data generation}
\label{sec:3}

\begin{figure*}[t]
	\centering
	\includegraphics[width=\textwidth]{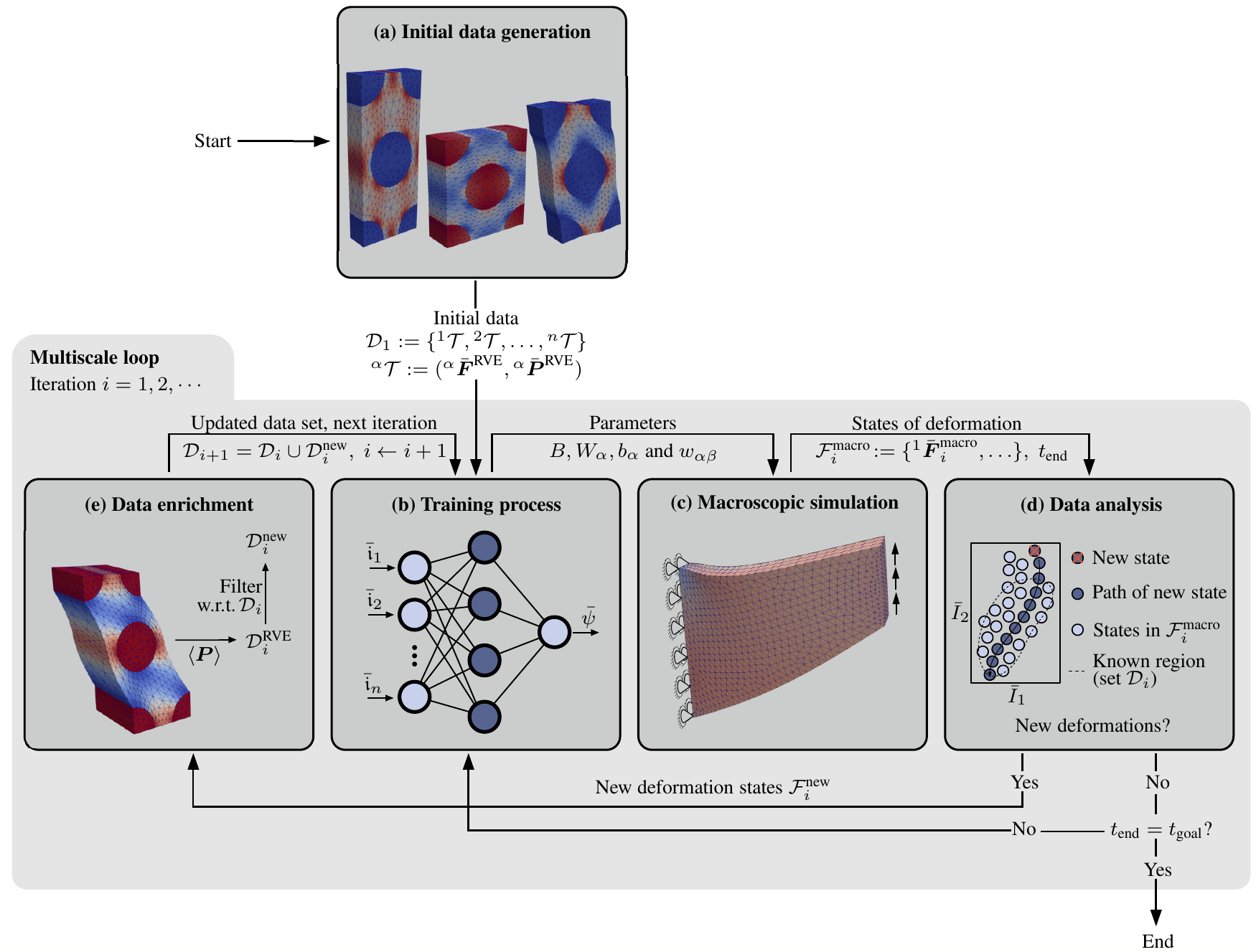}
	\caption{Schematic representation of the data-driven multiscale framework \emph{FE}${}^\textit{ANN}$: The framework starts with (a) initial data generation by homogenization. Afterwards, the steps (b) training process, (c) macroscopic simulation, (d) data analysis and (e) data enrichment are performed several times until no new data are added. Within the steps (d) and (e), the tolerances $\varepsilon_\text{tol,1}$ and $\varepsilon_\text{tol,2}$ are prescribed, respectively.}
	\label{fig:2}
\end{figure*}

\begin{algorithm}
	\begin{small}
	\caption{Procedure of the data-driven multiscale framework \emph{FE}${}^\textit{ANN}$ given as pseudo code.}
	\label{alg:1}	
	\begin{algorithmic}
		\State $\mathcal D_1 \leftarrow$ initial data generation ($\ind{1}\bte F,\ind{2}\bte F,\ldots,\ind{n}\bte F$)
		\For {$i=1,2,\ldots,n_\text{max}$}
		\Repeat 
		\State weights${}_i \leftarrow$ ANN training ($\mathcal D_i$)
		\State ($\mathcal F_i^\text{macro}, t_\text{end})\leftarrow$ macroscopic simulation (weights${}_i$)
		\State $\mathcal F_i^\text{new}\leftarrow$ detect unknown deformations ($\mathcal F_i^\text{macro},\mathcal D_i$,$\varepsilon_\text{tol,1}$)
		\Until {$t_\text{end}=t_\text{goal}$ \textbf{or}  $\mathcal F_i^\text{new}\ne \emptyset$}
		\If{$\mathcal F_i^\text{new} \ne \emptyset$}
		\State $\mathcal D_i^\text{RVE}\leftarrow$ homogenization($\mathcal F_i^\text{new}$)
		\State $\mathcal D_i^\text{new}\leftarrow$ filter($\mathcal D_i^\text{RVE}$,$\mathcal D_i$,$\varepsilon_\text{tol,2}$)
		\State $\mathcal D_{i+1} := \mathcal D_i \cup \mathcal D_i^\text{new}$
		\Else
		\State \textbf{break}
		\EndIf
		\EndFor
	\end{algorithmic}
	\end{small}
\end{algorithm}

Based on the summarized continuum theory, the following section introduces the proposed \emph{data-driven multiscale scheme FE}${}^\textit{ANN}$. 
The general procedure of this framework is basically subdivided into five steps referred as 
\begin{enumerate}
	\item[(a)] initial data generation,
	\item[(b)] training process of the ANN,
	\item[(c)] macroscopic simulation,
	\item[(d)] data analysis, as well as
	\item[(e)] data enrichment.
\end{enumerate}
After the framework is initiated with step (a), the steps (b)--(e) are executed within an \emph{overall loop} which is given in Algorithm~\ref{alg:1}. Accordingly, the ANN is trained with the current data set $\mathcal D_i$ of the iteration $i\in\N$, the macroscale problem is solved, unknown macroscopic states of deformation $\bte F$ are detected, and the macroscopic stresses $\bte P$ corresponding to the previously unknown deformations are generated via computational homogenization. The loop terminates in an iteration $i\ge 1$, after no further deformation states are found and the relevant space of deformation is thus completely sampled by $\mathcal D_{i}$. In order to prevent that the macroscopic simulation does not reach the final time step $t_\text{goal}$ -- which could be occur due to a failed convergence of the ANN -- but no new deformations are detected within $t_i \in \{t_0,t_1,\cdots,t_n = t_\text{end}<t_\text{goal}\}$, a further inner repeat loop is implemented. Therein, the steps (b)--(d) are performed multiple times, until the final time step $t_\text{goal}$ is reached or new states of deformation are found.
A schematic representation of the framework is given in Fig.~\ref{fig:2}. The single steps are described in detail in the following. 

In order to enable a fully automatized utilization of the framework, a Python wrapper which runs on a high performance cluster (HPC) using the \emph{Batch-System SLURM} (Simple Linux Utility for Resource Management) has been implemented. This wrapper processes the loop independently by starting the individual jobs and managing the results.

\subsection{Initial data generation (a)}
\label{Subsec:InitialData}
To start with, an initial data set \mbox{$\mathcal D_1:=\{\ind{1}\mathcal T,\ind{2}\mathcal T, \dots , \ind{n}\mathcal T\}$} consisting of $n\in\N$ data tuples \mbox{$\ind{\alpha}\mathcal T := (\ind{\alpha} \bte F^\text{RVE} , \ind{\alpha} \bte P^\text{RVE} ) \in  \L{2} \times \L{2}$} is generated. Basically, a low number of tuples is sufficient to initiate the multiscale scheme. Suitable states of deformation are simple load cases as, e.\,g., uniaxial tension and equibiaxial tension
\begin{align}
	\label{eq:uniaxial}
	[\bar F_{lK}^\text{RVE}] = \begin{bmatrix}
		\bar \lambda & 0 & 0\\
		0 & - & 0 \\
		0 & 0 & - 
	\end{bmatrix} \; , \;
	[\bar F_{lK}^\text{RVE}] = \begin{bmatrix}
		\bar \lambda & 0 & 0\\
		0 & \bar \lambda & 0 \\
		0 & 0 & - 
	\end{bmatrix}              
\end{align}
or simple shear
\begin{align}
	\label{eq:shear}
	[\bar F_{lK}^\text{RVE}] = \begin{bmatrix}
		1 & \bar \gamma & 0\\
		0 & 1 & 0 \\
		0 & 0 & 1 
	\end{bmatrix} \; ,            
\end{align}
where these load cases could be applied on the RVE in different directions.
In the equations above, $\bar \lambda \in \R_+$ and $\bar \gamma \in \R$ denote prescribed effective stretches and shears, respectively. The labeling with $(-)$ means that the corresponding coordinate of the effective 1st Piola-Kirchhoff stress $\bar P_{Kl}$ is prescribed to zero. The stresses $\ind{\alpha} \bte P^\text{RVE}$ belonging to the deformations $\ind{\alpha} \bte F^\text{RVE}$ are calculated from a \emph{computational homogenization} according to Subsect.~\ref{subsec:Homogenization}, whereby this is done by using an in house Matlab FE code. The periodic BCs are applied via the master node concept therein \cite{Haasemann2006}.

\subsection{ANN-based macroscopic surrogate model}
\label{SubSec:ANN-model}
Within the data-driven multiscale loop, an appropriate macroscopic surrogate model is needed. To this end, a physics-constrained ANN model, which a priori fulfills several physical conditions, is used, compare Subsect.~\ref{subsec:Hyper}. 

\begin{figure}
	\centering
	\includegraphics{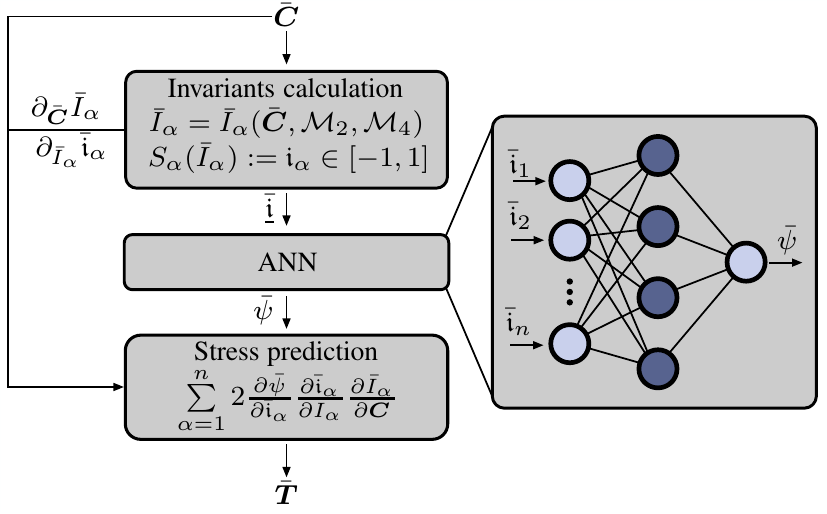}
	\caption{Structure of the macroscopic surrogate model based on a physics-constrained ANN. Note that $\mathfrak i_\alpha\in[-1,1]$ only holds for states included in the known region of training data.}
	\label{fig:3}
\end{figure}

Assuming that the macroscopic material symmetry group of the considered composite is known, and thus the structural tensors $\mathcal M_2$, $\mathcal M_4$ corresponding to the material are available, the invariants $\bar I_\alpha(\bte C,\mathcal M_2,\mathcal M_4)$ are determined first. These scalar values are mapped into a normalized domain, i.\,e., \mbox{$S_\alpha:\R \to [-1,1] \subset \R, \; \bar I_\alpha \mapsto \bar{\mathfrak i}_\alpha \forall \alpha \in\{1,2,\dots,n\}$} with respect to the training data set, and are arranged in a vector \mbox{$\bar{\underline{\mathfrak i}}=(\bar{\mathfrak i}_1,\bar{\mathfrak i}_2,\dots,\bar{\mathfrak i}_n) \in \R^{n\times 1}$}. Thereafter, the free energy is predicted by an FNN with the normalized invariants $\bar{\underline{\mathfrak i}}$ serving as input values and $\bar\psi$ as output. Restricting the network architecture to only one hidden layer containing $N$ neurons, it holds 
\begin{align}
	\label{eq:psi_ANN}
	\bar \psi^\text{ANN} := B + \sum_{\alpha=1}^{N} W_{\alpha}\,\Softplus\Big(\sum_{\beta=1}^{n} w_{\alpha\beta} \bar{\mathfrak i}_\beta + b_\alpha\Big) \; ,
\end{align}
where the monotonously increasing and convex \emph{Softplus} activation function \cite{Klein2021}
\begin{align}
	\Softplus: \R \to (0,\infty), \ x \mapsto \Softplus(x) := \ln(1 + \exp(x))
	\label{eq:softplus}
\end{align}
is used.\footnote{Note that a non-bounded activation function is necessary to fulfill the \emph{growth condition}, i.\,e., \mbox{$\bar \psi^\text{ANN}(\bte C)\to\infty$ as $(\bar J\to\infty \vee \bar J \to 0^+)$}. Besides the \emph{Softplus} activation function, further choices are possible, e.\,g., \emph{ELU} or \emph{ReLu}. However, in contrast to the \emph{Softplus} function, they are not twice continuously differentiable.} The stress prediction is finally done by applying Eq.~\eqref{eq:Comp}:
\begin{align}
	\bte T^\text{ANN} = \sum_{\alpha=1}^{n}2 \diffp{\bar \psi^\text{ANN}}{\bar{\mathfrak i}_\alpha} \diffp{\bar{\mathfrak i}_\alpha}{\bar I_\alpha}\diffp{\bar I_\alpha}{\bte C} \; . \label{eq:StressANN}
\end{align}
Thus, the surrogate model automatically fulfills several physical principles: \emph{thermodynamic consistency, objectivity} and \emph{material symmetry}. A graphical summary of the described surrogate model structure is given in Fig.~\ref{fig:3}.

In order to fulfill additional physical conditions, it may be necessary to incorporate further non-independent invariants $\bar I_\beta^*$ into the argument list of $\bar \psi^\text{ANN}$. In this case, Eqs.~\eqref{eq:psi_ANN} and \eqref{eq:StressANN} are to be modified according to
\begin{align}
	\bar \psi^\text{ANN} := B + \sum_{\alpha=1}^{N} W_{\alpha}  \Softplus\Big(\sum_{\beta=1}^{n} w_{\alpha\beta} \bar{\mathfrak i}_\beta  + \sum_{\beta \in \mathcal A} w^*_{\alpha\beta} \bar{\mathfrak i}^*_\beta +  b_\alpha\Big)
\end{align}
and
\begin{align}
	\bte T^\text{ANN} = \sum_{\alpha=1}^{n}2 \diffp{\bar \psi^\text{ANN}}{\bar{\mathfrak i}_\alpha} \diffp{\bar{\mathfrak i}_\alpha}{\bar I_\alpha}\diffp{\bar I_\alpha}{\bte C}
	+ \sum_{\beta\in\mathcal A}2 \diffp{\bar \psi^\text{ANN}}{\bar{\mathfrak i}_\beta^*} \diffp{\bar{\mathfrak i}_\beta^*}{\bar I_\beta^*}\diffp{\bar I_\beta^*}{\bte C}\; , \label{eq:StressANNmod}
\end{align}
where $\mathcal A := \{\gamma_1,\dots,\gamma_A\}, A = \vert \mathcal A\vert$, is a set containing the indices of the additional invariants $\bar I_\beta^*$. 
Accordingly, the \emph{growth condition} can be additionally guaranteed for $\bar J\to 0^+$ by including $\bar I_3^* := 1/\bar I_3$ as a further invariant and not only $\bar I_3$ itself, where $\bar I_3^*$ is defined to be independent, i.\,e., $\partial_{\bar I_3}\bar I_3^*=0$. Thus, $\mathcal A=\{3\}$ in this case. However, in addition, further \emph{constraints} have to be satisfied by the weights to ensure that the growth condition is fulfilled. As shown in Appendix~\ref{sec:App2}, a possible sufficient condition is given by
\begin{align}
	\begin{split}
		&\left(W_\alpha > 0 \forall \alpha \in \mathcal N\right) \cdots \\ 
		\cdots \wedge &\left(\exists \,w_{\alpha 3}>0 \text{ with } \alpha \in \mathcal N\right) \wedge \left(\exists \,w^*_{\alpha 3}>0 \text{ with } \alpha \in \mathcal N\right) \; ,
		\label{eq:conditionGrowth}
	\end{split}
\end{align}
Therein, the set $\mathcal N:=\{1,2,,\ldots,N\}$ contains the indices of the hidden layer neurons.
In the work \cite{Klein2021}, the growth condition is fulfilled in a similar way by adding an additional energy term which is not directly included in $\bar \psi^\text{ANN}$.

\subsubsection{Training process of the ANN (b)}
The ANN-based model is trained with respect to the current data set $\mathcal D_i$ in each iteration $i$ of the multiscale loop, where a random division into training and test data is made. 

Within the training process, the weights and bias values  $B, W_\alpha, b_\alpha$, $w_{\alpha\beta}$ and $w^*_{\alpha\beta}$ are then determined. 
In order to enable an adjustment of the ANN to the stress values, gradients of the output with respect to the input are inserted into the loss  
\begin{align}
	\mathcal L := \sum_\alpha \sqrt{(\ind{\alpha} \bar T^\text{ANN}_{11} - \ind{\alpha} \bar T_{11}^\text{RVE})^2 + \cdots + (\ind{\alpha} \bar T^\text{ANN}_{12} - \ind{\alpha} \bar T_{12}^\text{RVE})^2}
\end{align}
which is similar to \cite{Fernandez2020a,Linka2021,Vlassis2020}. The training is done by using the \emph{SLSQP optimizer} (Sequential Least Squares Programming). Thereby, the ANN is trained several times, where the parameters of the best achieved training state are stored at the end \cite{Kalina2021}.\footnote{Due to local minima within the loss function, the optimization procedure which is applied here depends on the starting values of the weights and biases. Thus, the network is trained several times to overcome this.}
An implementation of the described workflow is realized using \emph{Python, Tensorflow} and \emph{SciPy}. Within the training the constraint~\eqref{eq:conditionGrowth} can be switched on optionally.

Finally, in order to fulfill the \emph{normalization condition}, the bias value $B$ is chosen such that $\bar \psi^\text{ANN} = 0$ within the initial (undeformed) state.

\subsubsection{Implementation and macroscopic simulation (c)}
\label{Subsec:Macro}

The model equation \eqref{eq:psi_ANN} has been implemented within the FE toolbox \emph{FEniCS} \cite{Alnaes2015,Logg2012}. Therein, the stress relation given by Eq.~\eqref{eq:StressANN} and the material tangent
\begin{align}
	\bar{\tte C}^\text{ANN} &:= 4 \frac{\partial^2 \bar \psi^\text{ANN}}{\partial \bte C \partial \bte C} \in \L{4} \; ,
\end{align}
which is required within the solution via a standard Newton-Raphson scheme, are calculated by means of \emph{automatic differentiation}. 

In addition to weights and bias values of the trained ANN, problem specific structural tensors $\mathcal M_2^\text{macro}$, $\mathcal M_4^\text{macro}$ have to be prescribed within the macroscopic simulation, cf. Subsect.~\ref{subsubsec:MaterialSymmetry}. Note that the preferred directions on the macroscale does not necessarily have to match the ones of the RVE. A conversion which is necessary in this case is done in step (e), cf. Subsect.~\ref{Subsec:DateEnrichment}.

\subsection{Autonomous data mining}

Besides the physics-constrained surrogate model, the core feature of the data-driven multiscale framework is the data mining process which is done in a fully automatic manner.

\subsubsection{Data analysis (d)}
\label{Subsec:DataAnalysis}
Within each iteration of the overall multiscale loop, the local deformation states of the macroscopic body $\bar \B$ are collected. To this end, the deformation gradient $\bte F$ is stored at each quadrature point of the FE mesh and at each time increment \mbox{$t_j$}. Consequently, the body's full state of deformation is characterized by the set \mbox{$\mathcal F_i^\text{macro} :=\{\ind{1}\bte F^\text{macro}_i,\ind{2}\bte F^\text{macro}_i,\ldots,\ind{m}\bte F^\text{macro}_i\}$} which is a subset of $\L{2}$. 

Now, previously unknown deformations have to be detected, whereby it is meaningful to only consider states providing additional information for the material under observation. 
Since the intrinsic constitutive behavior of the material lives in the \emph{space of invariants} $\mathcal I \subset \R^{(n+A)\times 1}$, it is useful to perform a transformation into this space at this point \cite{Kalina2021}:
\begin{align}
	T: \L{2} \to \R^{(n+A)\times 1}, \bte F \mapsto (\bar I_1,\dots,\bar I_n, \bar I_{\gamma_1}^*,\dots,\bar I_{\gamma_A}^*)\; ,
	\label{eq:TransformAniso}
\end{align}
whereby $\mathcal M_2^\text{macro}$, $\mathcal M_4^\text{macro}$ and $\mathcal M_2^\text{RVE}$, $\mathcal M_4^\text{RVE}$ are used, respectively.
Thus, by making use of Eq.~\eqref{eq:TransformAniso}, the current data set $\mathcal D_i$ is compared to $\mathcal F_i^\text{macro}$ within the invariant space. If a state which is contained in \mbox{$\mathcal F_i^\text{macro}$} is unique within a given tolerance $\varepsilon_\text{tol,1}$, it is needed to enrich the data set for the next multiscale iteration step $i+1$. Thereby, the set $\mathcal F^\text{macro}_i$ is searched in an reverse manner, i.\,e., starting from the last time step $t_\text{end}$. If a unique state is identified in the step $t_n$, the full time series $\bte F(t_0), \bte F(t_1), \dots, \bte F(t_n)$ with $t_n \le t_\text{end}$ is added to  $\mathcal F_i^\text{new} \subset \L{2}$.\footnote{Note that it is useful to save the time series of a new deformation state, i.\,e., all states $\bte F(t_0,t_1,\dots,t_m)$ proceeding at a fixed quadrature point. This facilitates the application of the macroscopic deformation within the computational homogenization later on. Furthermore, an extension to path dependent materials requires this mandatory.} 
Thus, it is possible that deformation states in the set $\mathcal F_i^\text{new}$ are multiple with respect to $\mathcal F_i^\text{new}$ itself and/or $\mathcal D_i$. To avoid that this unnecessarily inflates the training process, a further filtering step is performed in the data enrichment step (e).

By using the described technique for the identification of relevant macroscopic deformation states, it is possible to reduce the number of time consuming microscale simulations following in the next step to a minimum.

\subsubsection{Data enrichment (e)}
\label{Subsec:DateEnrichment}

After $\mathcal F_i^\text{new}$ has been determined, it is necessary to identify the stresses belonging to the states $\ind{\alpha}\bte F^\text{macro}$. This is done by applying these states within the RVE simulations and calculating the stresses by volume averaging.
In order to use only one RVE and anyhow allow different \emph{microstructure orientations} in the macroscopic body, a transformation of $\ind{\alpha}\bte F$ is necessary before. To this end, the relation \mbox{$\bar \psi(\bte F) = \bar\psi(\te Q \cdot \bte F \cdot \te Q^\text{T})$} with $\te Q \in \Orth^+$, which holds for arbitrary material symmetry groups, is used. Thus, the collected state of deformation $\ind{\alpha}\bte F^\text{macro}$ from the macroscopic sample is processed by the  \emph{Euclidean transformation} 
\begin{align}
	\ind{\alpha}\bte F^\text{RVE} = \te Q \cdot \ind{\alpha}\bte F^\text{macro} \cdot \te Q^\text{T} \; .
	\label{eq:Rotation}
\end{align} 
The tuples \mbox{$\ind{\alpha}\mathcal T = (\ind{\alpha} \bte F^\text{RVE} , \ind{\alpha} \bte P^\text{RVE} )$} consisting of applied deformations and corresponding stresses are then collected in the set $\mathcal D_i^\text{RVE}$.

As already mentioned above, it is possible that data tuples in the set $\mathcal D_i^\text{RVE}$ are multiple with respect to $\mathcal D_i^\text{RVE}$ itself and/or $\mathcal D_i$. Thus, a further filtering step is performed, where multiple tuples are sorted out with respect to $\mathcal D_{i}$ and $\mathcal D_i^\text{RVE}$ within a given tolerance $\varepsilon_\text{tol,2}$. To this end, a transformation to the invariant space is used again, see Subsect.~\ref{Subsec:DataAnalysis}.
Finally, the enriched data set for the next iteration of the multiscale loop follows from \mbox{$\mathcal D_{i+1} = \mathcal D_{i} \cup \mathcal D_{i}^\text{new}$}, where $\mathcal D_{i}^\text{new}$ only contains relevant and unique tuples of the added deformations and corresponding stresses.

\section{Examples}
\label{sec:4}
In order to demonstrate the ability of the developed \emph{data-driven multiscale approach FE}${}^\textit{ANN}$ described in Sect.~\ref{sec:3}, it is applied to the solution of three numerical examples within this section. Specifically, three macroscopic structures -- a cuboid with a circular hole, a torsional sample and the Cook membrane -- are considered. All of them consist of a \emph{fiber reinforced composite} revealing a highly nonlinear behavior of the individual phases. 

\subsection{Microscopic properties of the composite}
\subsubsection{Constitutive behavior}
\label{subSec:Const}

\begin{table*}
	\caption{Material parameters for matrix and fiber phases of the composite described by the Ogden model~\eqref{eq:OgdenComp}. Initial shear modulus $G^\text{init}$ and Poisson's ratio $\nu^\text{init}$ as well as parameter sets $\mu_p$, $\alpha_p$ and $\kappa$. The parameters of the matrix phase are chosen according to Kalina~et~al.~\cite{Kalina2021}.}
	\centering
	\footnotesize
	\begin{tabular}{c c c c c c c c c c}
		& $G^\text{init} \; / \; \si{\kilo\pascal}$ & 	$\nu^\text{init} \; / \; -$ &
		$\mu_1 \; / \;  \si{\kilo\pascal}$ & $\mu_2 \; / \;  \si{\kilo\pascal}$ & $\mu_3 \; / \;  \si{\kilo\pascal} $ & $\alpha_1 \; / \; -$ & $\alpha_2 \; / \; -$ & $\alpha_3 \; / \; - $ & $\kappa \; / \;  \si{\kilo\pascal}$\\
		\hline
		Matrix & $100.0$ & $0.44$ & $-26.62$ & $29.04$ & $0.0098$ & $-5.0$ & $2.3$ & $12.0$ & $800.0$ \\
		Fibers & $1000.0$ & $0.40$ & $1000.0$ & $-$ & $-$ & $2.0$ & $-$ & $-$ & $4666.7$
	\end{tabular}
	\label{tab:ConstPara}
\end{table*}

The constitutive behavior of the composite's individual components, i.\,e., fibers and matrix, is described by a hyperelastic \emph{Ogden model} \cite{Ogden1997}. It is given by the free energy density function 
\begin{align}
	\psi := \sum_{p=1}^{N_\text{O}} \frac{\mu_p}{\alpha_p}
	\left[
	\sum_{\beta=1}^{N_\lambda} \nu_\beta \left(\lambda_\beta^\text{iso}\right)^{\alpha_p} -3 \right]
	+ \frac{\kappa}{4} \left(J^2 - 2 \ln J -1 \right) \; ,
\end{align}
where, $\mu_p,\alpha_p \in \R$ and $\kappa \in \R_+$ are model parameters. The symbol $\R_+ \ni \lambda_\beta^\text{iso} := J^{-1/3} \lambda_\beta$ denotes the isochoric principal stretches following from the Flory split \cite{Flory1961}:  
\begin{align}
	\te F = \te F^\text{vol} \cdot \te F^\text{iso} \; \text{with} \; \te F^\text{iso}=J^{-1/3} \te F \; \text{and} \; \det \te F^\text{iso} \equiv 1 \; ,
	\label{eq:Flory}
\end{align}
i.\,e., the multiplicative decomposition of $\te F$  into volumetric $\te F^\text{vol}$ and isochoric $\te F^\text{iso}$ parts. Note that the introduced material parameters are related to initial shear modulus $G^\text{init}$ and Poisson's ratio $\nu^\text{init}$ via the relations
\begin{align}
	G^\text{init}=\frac{1}{2} \sum_{p=1}^{N_\text{O}} \alpha_p \mu_p \quad \text{and} \quad
	\kappa = \frac{2}{3} G^\text{init} \frac{1+\nu^\text{init}}{1-2\nu^\text{init}} \; ,
\end{align}
respectively. Furthermore, in order to  guarantee a physically meaningful behavior, the parameters $\alpha_p$ and $\mu_p$ have to be restricted by the following constraints: \mbox{$(\alpha_p < -1 \, \vee \, \alpha_p\ge 2$} and \mbox{$\mu_p \alpha_{p} > 0) \, \forall p \in \{1,2,\cdots N_\text{O}\}$} \cite{Ogden1997}. The stress of the Ogden model is determined by using Eq.~\eqref{eq:consistent} and follows to
\begin{align}
	\label{eq:OgdenComp}
	\begin{split}
		\te T = \sum_{\beta=1}^{N_\lambda} 
		\Biggl[
		&\frac{1}{\lambda_\beta^2} \sum_{p=1}^{N_\text{O}} \mu_p \left(
		(\lambda_\beta^\text{iso})^{\alpha_p} -\frac{1}{3} \sum_{\gamma=1}^{N_\lambda}
		\nu_\gamma(\lambda_\gamma^\text{iso})^{\alpha_p}\right) 
		\\
		&+ \frac{\kappa}{2} \lambda_\beta^{-2} (J^2-1)
		\Biggr] \te P^\beta \; .
	\end{split}
\end{align}
The calculation of the projection tensors $\te P^\beta$ related to the right Cauchy-Green deformation tensor $\te C$ is given in Eq.~\eqref{eq:projection}.

Within the following numerical examples, the material parameters given in Tab.~\ref{tab:ConstPara} are chosen.	With this choice, a highly nonlinear behavior of the matrix phase is achieved, cf. \cite{Kalina2021}. The parameters of the fibers are chosen in such a way, that a \emph{neo-Hookean model} results.
Note that the selected parameters are not related to a real material. 

\subsubsection{Microstructure and RVE definition}

\begin{figure}
	\centering
	\includegraphics{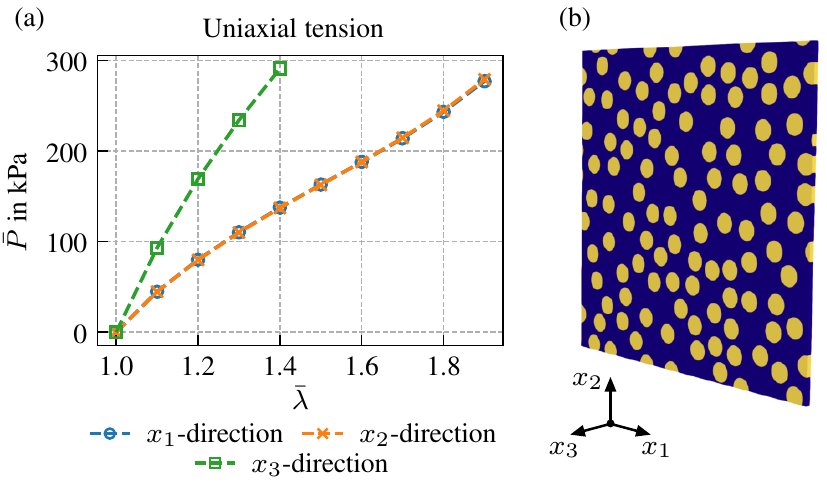}
	\caption{Fiber reinforced composite: (a) effective stress-stretch curves for uniaxial tension in different loading directions and (b) periodic RVE with 100 fibers and $\phi=\SI{30}{\percent}$ volume fraction. The size of the RVE is given by $(323.6 \times 323.6 \times 10)\,\si{\micro\meter}$ and the fiber radius is $\SI{10}{\micro\meter}$.}
	\label{fig:4}
\end{figure}

Besides the constitutive behavior of the individual components, the geometric arrangement of the individual phases has to be defined. In order to choose a realistic microstructure, it is characterized by a \emph{monodisperse stochastic arrangement} of fibers with $\phi=\SI{30}{\percent}$ volume fraction here.\footnote{It should be noted, that it is not sufficient to represent the microstructure of a fiber reinforced composite with an overall transversely isotropic behavior by a hexagonal unit cell if finite strains are considered. This is due to the loss of the material symmetry which results from the
	deformation of the RVE cell, cf. Appendix~\ref{sec:App1}.} Furthermore, a minimum distance of $d:=0.4 R$ with respect to the fiber radius $R=\SI{10}{\micro\meter}$ is chosen.  

According to the homogenization concept, a suitable RVE which is sufficiently large to capture the essential statistic properties of the microstructure is needed. Herein, this condition is checked by applying the $\chi^2$-test proposed by Gittman~et~al.~\cite{Gitman2007}. Thereby, $n$ randomly generated RVEs with fixed number of inclusions and volume fraction are generated. The effective response of these RVEs is then determined by a computational homogenization, whereby a specific load case is chosen depending on the property $a$ of interest, e.\,g., this could be a stress or a stiffness component. Based on these results, a \emph{statistical analysis} is done by evaluating the scatter of $a$ by means of the quantity
\begin{align}
	\chi^2:=\sum_{i=1}^n\frac{\left( a_i - \langle a \rangle_n  \right)^2}{\langle a \rangle_n} \; \text{with} \; \langle a\rangle_{n} &:=\frac{1}{n} \sum_{i=1}^{n} a_i \; .
	\label{eq:chiSquare}
\end{align}
If the accuracy of $a$ is sufficient, i.\,e., $\chi^2\le \varepsilon_\text{tol}$, the tested sample size is the final RVE size. Otherwise, the sample size have to be increased and the analysis will be repeated.

The described $\chi^2$-test has been applied to the fiber reinforced material. Thereby, a tolerance of $\varepsilon_\text{tol}=\SI{2.5}{\percent}$ and a number of $n=10$ RVEs for each RVE-class with $N^\text{inc}$ inclusions have been chosen. As representative load cases, uniaxial tensions according to Eq.~\eqref{eq:uniaxial}${}_1$ are applied into the $x_1$-, $x_2$- and the $x_3$-direction, i.\,e., perpendicular and parallel to the fiber orientation.
Thereby, the stress into the tension direction is evaluated according to Eq.~\eqref{eq:chiSquare}. The prescribed tolerance is finally reached for the RVE-class with $N^\text{inc}=100$ fibers. 
The determined stress-stretch curves and the chosen RVE are depicted in Fig.~\ref{fig:4} for the final RVE. Accordingly, a highly nonlinear and anisotropic effective response occurs. 

Generation and meshing of the periodic cells have been done by using the python tool \emph{gmshModel}\footnote{The python tool gmshModel is freely accessible under \url{https://gmshmodel.readthedocs.io/en/latest/}.}, whereby the placement of the fibers is realized via the random sequential adsorption (RSA) algorithm \cite{Torquato2002}. The RVEs were meshed by tetrahedron elements with 10 nodes. A total of 94,456 elements is reached for the final RVE with 100 fibers.

\begin{figure*}
	\centering
	\includegraphics{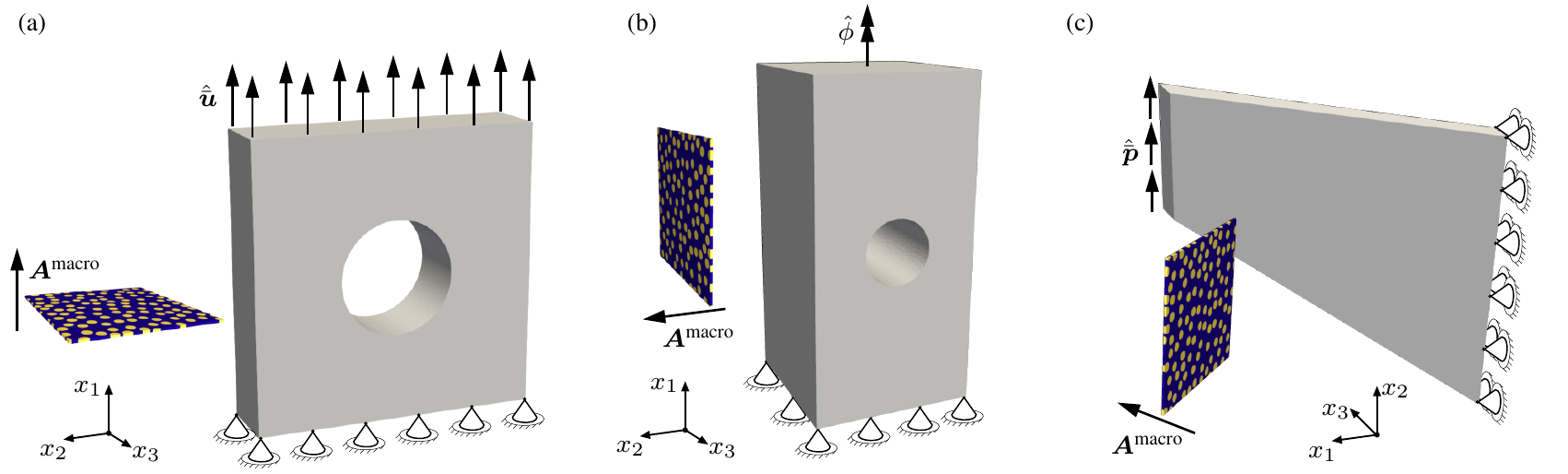}
	\caption{Macroscopic geometries with applied BCs and fiber direction $\ve A^\text{macro}$: (a) cuboid under tension with $\ve A^\text{macro}=\ve e_1$ and $\hat{\bve u}=\SI{40}{\milli\meter}$, (b) torsional sample with $\ve A^\text{macro}=\ve e_2$ and $\hat{\bar \phi}=45^\circ$ as well as (c) Cook's membrane with $\ve A^\text{macro}=(\ve e_1 + \ve e_3)/\sqrt{2}$ and $|\hat{\bve p}|=\SI{0.5}{\kilo\pascal}$.}
	\label{fig:5}
\end{figure*}

\subsection{Application of the data-driven multiscale framework for the simulation of macroscopic samples}

After the definition of the composite's microscopic properties, the data-driven framework \emph{FE}${}^\textit{ANN}$ is applied for the simulation of several multiscale problems. The macroscopic geometries and applied BCs are depicted in Fig.~\ref{fig:5}.

In all examples, the following specifications and meta parameters have been used:
The effective \emph{transversely isotropic} behavior, which results from the RVE homogenization, is described by using the set \mbox{$\bar{\V I}^\shortparallel:=(\bar I_1,\bar I_2,\bar I_3,\bar I_4, \bar I_5,\bar I_3^*)\in \R^{6\times 1}$} consisting of six invariants, cf. Subsect.~\ref{SubSubSec:AnisotropyClass}. As discussed in Subsect.~\ref{SubSec:ANN-model}, the non-independent invariant $\bar I_3^*=1/\bar I_3$ is added to allow the \emph{growth condition} to be satisfied by construction of the network architecture. 
As already mentioned, this requires additional constraints for the weights, e.\,g., Eq.~\eqref{eq:conditionGrowth}. However, training under these constraints results in noticeably worse stress predictions. Thus, in order to achieve maximum prediction quality, the constraints were not considered within the training here. The adapted architecture is nevertheless used to allow a better comparability to the same network architecture trained with the constraint, see the study in  Appendix~\ref{sec:App2}.
The ANNs used as the surrogate model consist of only \emph{one hidden layer} with 15 neurons which has been shown to be sufficiently accurate, where the networks are trained 25 times in each macroscopic iteration. For comparison see the study given in \cite{Kalina2021}. 

As shown in Fig.~\ref{fig:4}(b), the fiber orientation of the RVE is given by $\ve A^\text{RVE}=\ve e_3$, where $\ve e_3$ is the Cartesian base vector. In order to transform deformation states for the general case $\ve A^\text{macro} \ne \ve A^\text{RVE}$ by using Eq.~\eqref{eq:Rotation}, \emph{Rodrigues' rotation formula} 
\begin{align}
	\te Q = \ve N \otimes \ve N + \cos (\alpha) \left[\te 1 - \ve N \otimes \ve N\right] + \sin(\alpha) \ve N \times \te 1	\; .
\end{align}
is applied to determine $\te Q$. Therein, $\ve N := \ve A^\text{macro} \times \ve A^\text{RVE}$ is the unit normal vector and  \mbox{$\alpha := \measuredangle(\ve A^\text{macro}, \ve A^\text{RVE})$} the angle between the fiber directions.

Finally, the tolerance for the detection of unique deformation states within the invariant space $\mathcal I^\shortparallel$ is set to $\varepsilon_\text{tol,1}=\SI{5}{\percent}$. The tolerance for the filtering step is set to $\varepsilon_\text{tol,2}=\SI{1}{\percent}$, cf. Subsects.~\ref{Subsec:DataAnalysis} and \ref{Subsec:DateEnrichment}.

\subsubsection{Cuboid under tension}

\begin{figure*}
	\centering
	\includegraphics{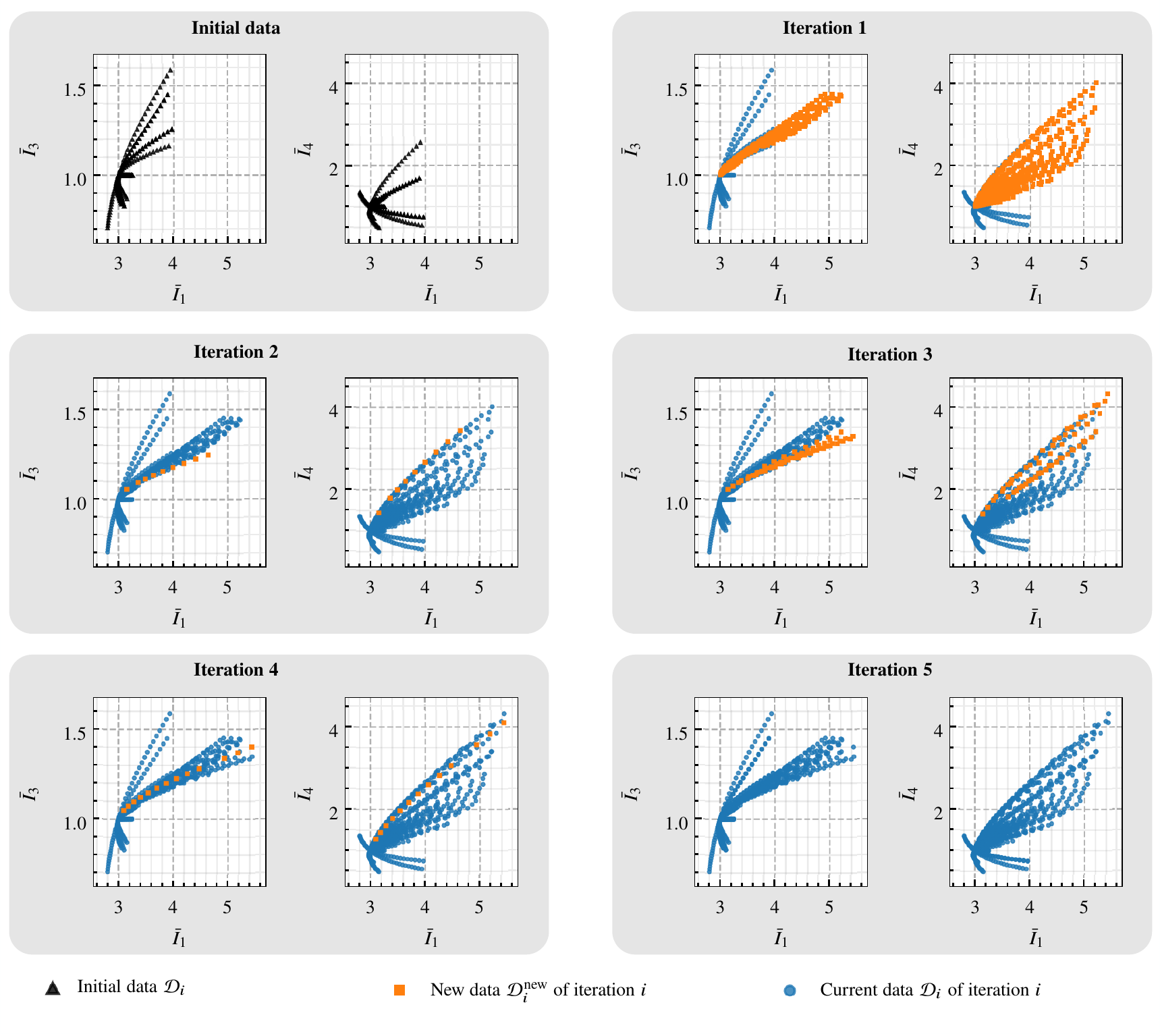}
	\caption{Autonomous sampling of the invariant space $\mathcal I^\shortparallel$ within the data-driven multiscale loop applied to the simulation of the cuboid under tension. Exemplarily, the sectional planes $\bar I_1$-$\bar I_3$ and $\bar I_1$-$\bar I_4$ are depicted. Starting with initial data set, relevant states are detected within the space $\mathcal I^\shortparallel$.}
	\label{fig:6}
\end{figure*}

As a first example, a cuboid with a circular hole is loaded by tension. To this end, a displacement of $\hat{\bve u} = 0.4 \bar L_{x_1}$ is prescribed at the top surface, where $\bar L_{x_1}$ is the initial length in $x_1$-direction. The fiber orientation is chosen to $\ve A^\text{macro} = \ve e_1$. The cuboid's geometric dimensions are specified by $\bar L_{x_1}\times\bar L_{x_2} \times \bar L_{x_3} = (100\times 100 \times 25) \, \si{\milli\meter}$. The hole in the center has a radius of $\SI{30}{\milli\meter}$.

\paragraph{Multiscale iterations}
The single iterations within the loop are described for the multiscale simulation of the cuboid in the following. 

\begin{table}
	\centering
	\caption{Load cases for the generation of the initial data. The effective deformations $\bte F^\text{RVE}$ are prescribed according to Eqs.~\eqref{eq:uniaxial} and \eqref{eq:shear}.}
	\begin{tabular}{p{2cm} c c c p{2.5cm}}
		Load case & $\bar \lambda_\text{max}$ & $\bar \lambda_\text{min}$ & $\bar \gamma_\text{max}$  & directions\\
		\hline 
		Uniaxial tension & $1.60$ & $-$ & $-$ & $x_1$,$x_2$,$x_3$\\
		Equibiaxial tension & $1.30$ & $-$ & $-$ & $x_1$-$x_2$,$x_1$-$x_3$,$x_2$-$x_3$\\
		Uniaxial compression & $-$ & $0.70$ & $-$ & $x_1$,$x_2$,$x_3$\\
		Equibiaxial compression & $-$ & $0.85$ & $-$ & $x_1$-$x_2$,$x_1$-$x_3$,$x_2$-$x_3$\\
		Simple shear & $-$ & $-$ & $0.50$ & $x_1$-$x_2$,$x_1$-$x_3$,$x_2$-$x_3$, \\
		&  &  & & $x_2$-$x_1$,$x_3$-$x_1$,$x_3$-$x_2$
	\end{tabular}
	\label{tab:2}
\end{table}

To start the algorithm, the initial data have to be generated first, cf. Subsect.~\ref{Subsec:InitialData} step (a). Here, a total number of 18 load cases -- six uniaxial tension and compression, six equibiaxial tension and compression as well as six simple shear states -- are prescribed to the fiber reinforced RVE. Different loading directions are considered to collect knowledge about the composite's overall anisotropy. The applied stretches and shears as well as the directions are given in Tab.~\ref{tab:2}. In order to avoid that data providing the same physical information are contained multiple times, a filtering process is applied in the invariant space $\mathcal I^\shortparallel$.\footnote{Due to the material symmetry (transverse isotropy), several loadings are equivalent from the point of the material, e.\,g., uniaxial tension in $x_1$- or $x_2$-direction. Note that the fiber direction of the RVE is $\ve A^\text{RVE}=\ve e_3$.}
Within $\mathcal I^\shortparallel$, the collected states cover only a sparse region which is pervaded by a few paths consisting of 193 tuples $\ind{\alpha}\mathcal T$. This is shown in Fig.~\ref{fig:6}, where the set is exemplarily visualized in the sectional planes $\bar I_1$-$\bar I_3$ and $\bar I_1$-$\bar I_5$. However, due to the physical knowledge which is incorporated into the ANN-based macroscopic surrogate model, this sparse data set is sufficient to initiate the multiscale scheme. 

Now, the algorithm enters the multiscale loop with iteration 1 and the initial data set is labeled as $\mathcal D_1$. It is used to train the \emph{physics-constrained ANN} which is afterwards utilized as the RVE surrogate model within the macroscopic simulation, cf. Subsect.~\ref{SubSec:ANN-model} steps (b) and (c). Note that the prediction of the sample's macroscopic fields, i.\,e., displacement $\bve u(\ve X,t_i)$, deformation $\bte F(\ve X,t_i)$, stress $\bte P(\ve X,t_i)$, etc., is only an initial guess in this first iteration. This is due to the fact that the ANN does not know the occurring stress-deformation states for the most part at this point. However, it is still sufficient to identify new deformation states. As described in Subsect.~\ref{Subsec:DataAnalysis}, this is done by transforming the sample's deformation states, collected into the set $\mathcal F^\text{macro}_1$, into the invariant space $\mathcal I^\shortparallel$ and to identify unique states by a comparison to $\mathcal D_1$. As already mentioned, the full time series $\bte F(t_0), \bte F(t_1), \dots, \bte F(t_n)$ with $t_n \le t_\text{end}$ is added to  $\mathcal F_i^\text{new}$ if a unique state is identified in the step $t_n$. These deformation paths are prescribed in the RVE simulation and the corresponding stresses $\bte P$ are determined. After the subsequent filtering process with the tolerance $\varepsilon_\text{tol,2}$, a total number of 495 unique tuples is identified. They are given in Fig.~\ref{fig:6}. 

\begin{figure*}
	\centering
	\includegraphics{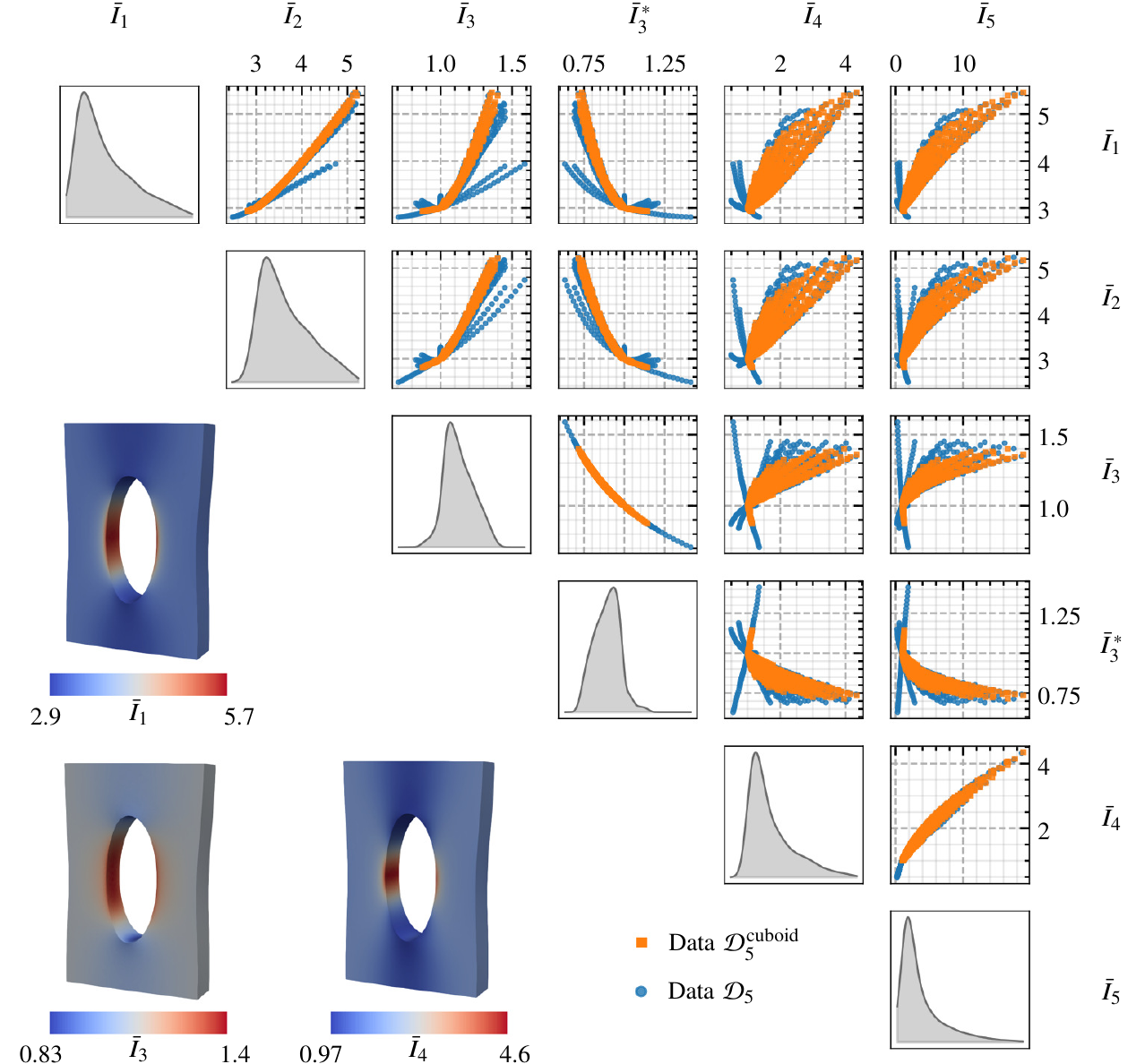}
	\caption{Invariant space $\mathcal I^\shortparallel$ of the cuboid under tension within the last multiscale iteration 5: (a) sectional planes with the full data set $\mathcal D_5$ and the final prediction $\mathcal D_5^\text{cuboid}$ of the cuboid's deformation mapped into $\mathcal I^\shortparallel$ as well as (b)--(d) field distribution of the invariants $\bar I_1$, $\bar I_3$ and $\bar I_4$ on the deformed sample, respectively}
	\label{fig:7}
\end{figure*}

\begin{table}
	\centering
	\caption{Multiscale iterations of the cuboid under tension: Tuples in the current data set $\mathcal D_i$ and the new states collected in $\mathcal D_i^\text{new}$ as well as reached time step $t_\text{end}$ within the macro simulation.}
	\begin{tabular}{c c c c}
		Iteration $i$ & Tuples in $\mathcal D_i$ & Tuples in $\mathcal D_i^\text{new}$ & $t_\text{end}/t_\text{goal}$\\
		\hline 
		1 & 193 & 495 & 1 \\
		2 & 688 & 9 & 12/15 \\
		3 & 697 & 60 & 1 \\
		4 & 757 & 13 & 1 \\
		5 & 770 & 0 & 1
	\end{tabular}
	\label{tab:3}
\end{table}

In iteration 2, the new information are used to enrich the data set $\mathcal D_2=\mathcal D_1\cup\mathcal D_1^\text{new}$. This set is used to train the ANN which is then applied for the prediction of the sample's macroscopic fields. As shown in Fig.~\ref{fig:6}, only a low number of new tuples is now detected from $\mathcal F_2^\text{macro}$. 
This is due to the fact that the macroscopic simulation terminates already in time step $t_\text{end} = t_{12}<t_\text{goal}$, cf. Tab.~\ref{tab:3}.

However, using these new data, the prediction of the macroscopic simulation in the subsequent 3rd iteration yields that the sample's deformation field maps to a new region within $\mathcal I^\shortparallel$ which consists of 60 tuples.  
The described process -- consisting of the steps training process, macroscopic simulation, data analysis and data enrichment -- is repeated until no further unique data are found which is the case in iteration 5. Note that an overlap of tuples from $\mathcal D_i$ and $\mathcal D_i^\text{new}$ is possible, since a state is already unique if it deviates only in one of the six relevant invariants. Thus, an intersection in the depicted sectional planes may occur.

In advance, the macroscopic sample's states $\mathcal F_5^\text{macro}\to\mathcal D_5^\text{macro}$ mapped to $\mathcal I^\shortparallel$ and the data set $\mathcal D_5$ of iteration 5 are depicted within the full invariant space in Fig.~\ref{fig:7}. It becomes apparent, that $\mathcal D_5 \setminus \mathcal D_1$ contains tuples which are not included into $\mathcal F_5^\text{macro}$. Thus, as described above, the predictions of iteration 1--4 are only necessary to gradually approach the correct result of the macroscopic simulation. However, due to the fact that the collected deformations are prescribed within the RVE simulations to determine the corresponding stresses, no defective information could enter the data set. 

\begin{figure}
	\centering
	\includegraphics{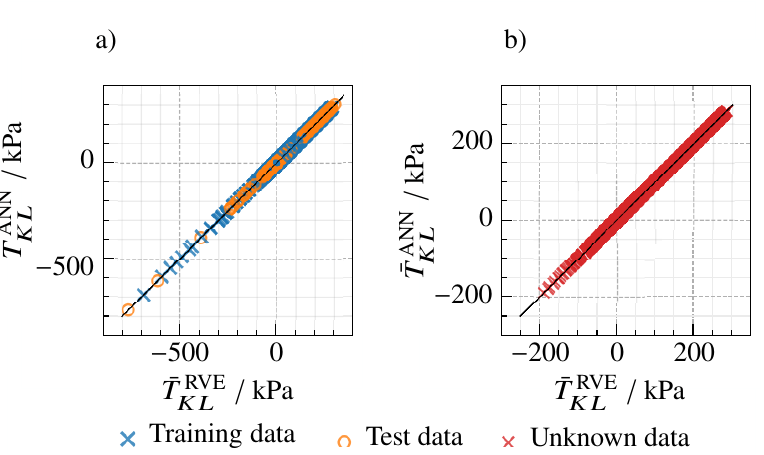}
	\caption{Predictions of the ANN compared to reference stresses obtained from RVE simulation: (a) data $\mathcal D_5$ divided into training and test data for the ANN training process and (b) completely unknown data (within the deformation space).}
	\label{fig:8}
\end{figure}

\begin{figure}
	\centering
	\includegraphics{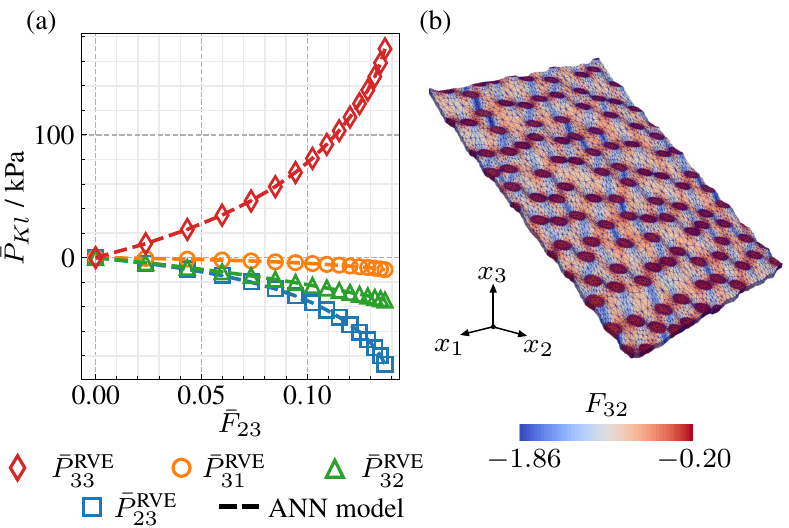}
	\caption{Comparison of the homogenized RVE response and the ANN-prediction: (a) Full loading path and (b) deformed RVE with local deformation field corresponding to the final time step $t_\text{goal}$.}
	\label{fig:10}
\end{figure}

The stress predictions of the ANN for all tuples in $\mathcal D_5$ are compared to the reference values of the RVE simulations in Fig.~\ref{fig:8}(a). As shown there, an almost perfect prediction occurs for both, training and test data. A single loading path and the corresponding deformed RVE at time $t_\text{goal}$ are depicted in Fig.~\ref{fig:10} exemplarily.

\paragraph{Validation}

\begin{figure}
	\centering
	\includegraphics[width=\columnwidth]{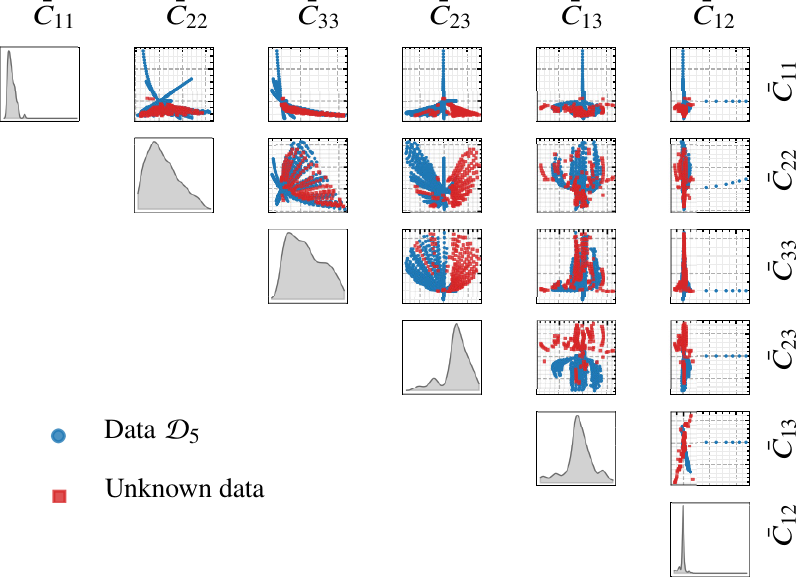}
	\caption{Deformations occurring in the cuboid simulation: data set $\mathcal D_5$ mapped into the $\bte C$-space (final iteration) and completely unknown data. The fiber direction is given by $\ve A^\text{RVE} = \ve e_3$.}
	\label{fig:9}
\end{figure}

In order to demonstrate the quality of the developed \emph{FE}${}^\textit{ANN}$ approach, the full deformation space of the cuboid is analyzed and the single states are applied to the RVE simulation, respectively. To this end, the symmetric right Cauchy-Green deformation tensors $\ind{\alpha}\bte C$ are calculated from $\mathcal F_5^\text{macro}$ and included into the new set 
\begin{align}
	\mathcal C_5^\text{macro}=\{\ind{1}\bte C^\text{macro}, \ind{2}\bte C^\text{macro}, \dots, \ind{n}\bte C^\text{macro}\}\subset \Sym \; .
\end{align}
This set is then compared to the set $\mathcal D_5$ within the space of the tensor $\bte C\in\Sym$. Thereby, the contained deformation gradients $\ind{\alpha}\bte F^\text{RVE}$ in $\mathcal D_5$ are rotated back to $\ind{\alpha}\bte F^\text{macro}$ by inverting Eq.~\eqref{eq:Rotation}. Subsequently, they are transformed to $\bte C$-values to enable a comparison with $\mathcal C_5^\text{macro}$. The unique states are then determined within a tolerance of $\SI{5}{\percent}$. The states contained in $\mathcal D_5$ and the unique, unknown states in the deformation space, i.\,e., $\mathcal C_5^\text{macro}\setminus\mathcal D_5$, are depicted in Fig.~\ref{fig:9}. Accordingly, a wide region of $\mathcal C_5^\text{macro}$ do not intersect with the training data set which results from the transformation into the invariant space during the multiscale loop. 

The unknown states are prescribed within RVE simulations to get the corresponding stress values. These stresses are then compared to the predictions of the ANN which has been trained by $\mathcal D_5$. As shown in Fig.~\ref{fig:8}(b), an almost perfect prediction is observed also for these states. Consequently, the proposed \emph{FE}${}^\textit{ANN}$ approach has shown to be highly accurate. 

\begin{figure}
	\centering
	\includegraphics{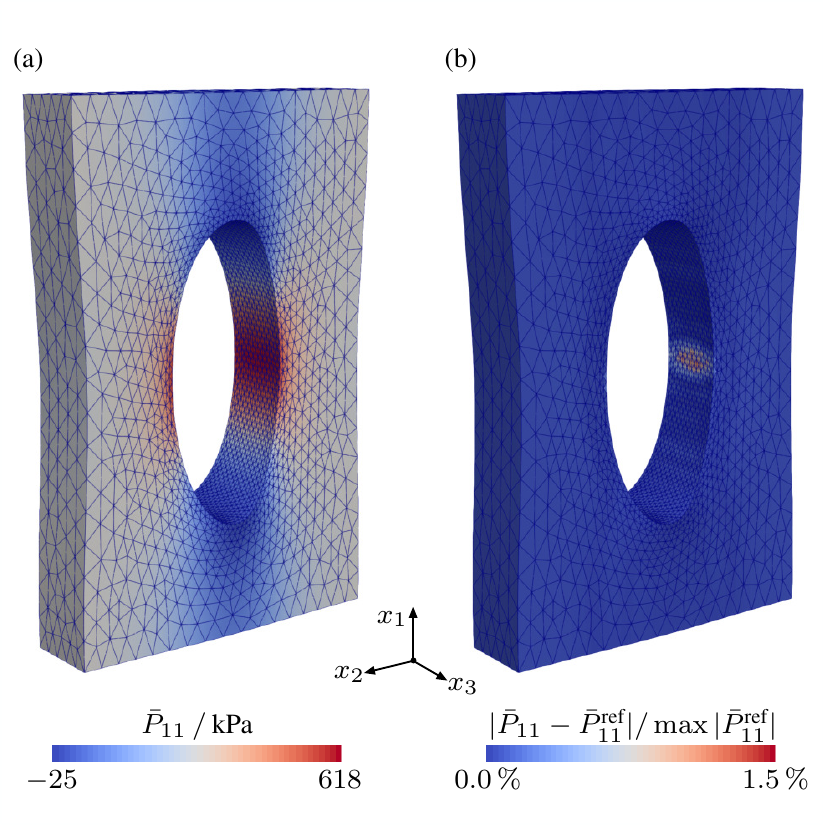}
	\caption{Stress field in the cuboid under tension: (a) macroscopic stress field $\bar P_{11}$ on the deformed configuration and (b) relative error of the determined stress field $\bar P_{11}$ with respect to $\bar P_{11}^\text{ref}$ obtained from a solution of the multiscale scheme with a reduced detection tolerance of $\varepsilon_\text{tol,1}^\text{ref}=2.5\,\%$. Within the original simulation, the detection tolerance was $\varepsilon_\text{tol,1}=5\,\%$.}
	\label{fig:11}
\end{figure}

Furthermore, in order to demonstrate that the chosen tolerance $\varepsilon_\text{tol,1}=\SI{5}{\percent}$ within the multiscale loop is sufficient, the achieved macroscopic solution is compared to a solution in which this tolerance is prescribed to $\SI{2.5}{\percent}$.\footnote{In want of a fully coupled FE${}^2$ implementation, this comparison is used.} 
As shown in Fig.~\ref{fig:11}, a relative error of below $\SI{2}{\percent}$ with respect to the reference solution occurs for the stress $\bar P_{11}$

\subsubsection{Torsional sample}

\begin{figure}
	\centering
	\includegraphics{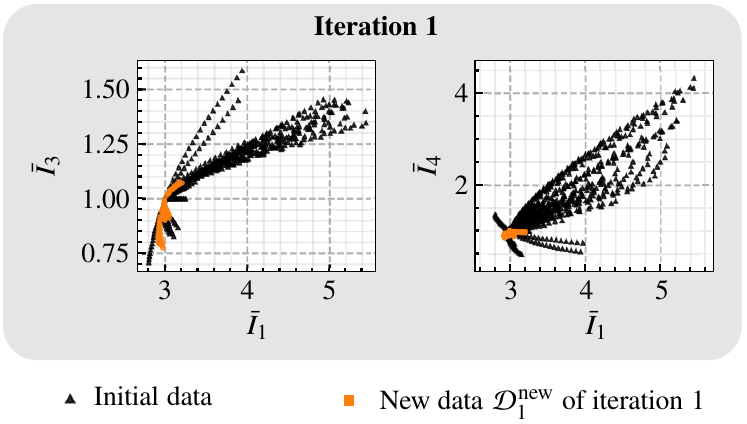}
	\caption{Sampling of the invariant space $\mathcal I^\shortparallel$ within the simulation of the torsional sample. The initial data set contains the load cases given in Tab.~\ref{tab:2} and the data collected from the cuboid simulation. The multiscale scheme was terminated in iteration 2. Exemplarily, the sectional planes $\bar I_1$-$\bar I_3$ and $\bar I_1$-$\bar I_4$ are depicted.}
	\label{fig:12}
\end{figure}

\begin{figure}
	\centering
	\includegraphics{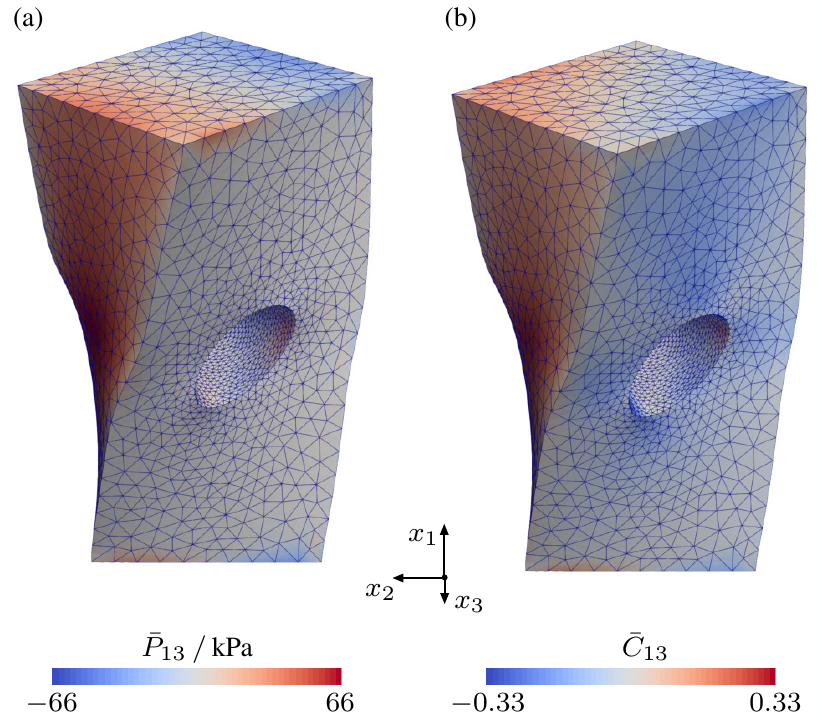}
	\caption{Torsional sample: distribution of the macroscopic 1st Piola-Kirchhoff stress $\bar P_{13}$ and right Cauchy-Green deformation $\bar C_{13}$ at the final time step. The surface plots are given on the deformed configuration $\bar{\mathcal B}$.}
	\label{fig:13}
\end{figure}

After the detailed description of the multiscale loop within the solution process of the cuboid geometry, a further example is considered. Now, the \emph{FE}${}^\textit{ANN}$ scheme is applied to the solution of a \emph{torsional sample} with a circular hole given in Fig.~\ref{fig:5}(b). Accordingly, the fiber direction of the macro sample is chosen as $\ve A^\text{macro}=\ve e_2$ in this example. Nevertheless, due to the applied transformation given by Eq.~\eqref{eq:Rotation}, the same RVE with $\ve A^\text{RVE}=\ve e_3$ is used. The torsional sample is loaded by specifying a distortion of $\hat{\bar \phi} = 45^\circ$ around the $x_1$-axis.
The sample's geometric dimensions are specified by $\bar L_{x_1}\times\bar L_{x_2} \times \bar L_{x_3} = (200\times 100 \times 100) \, \si{\milli\meter}$. The hole in the center has a radius of $\SI{40}{\milli\meter}$.

In order to minimize the computational effort, the scheme is initiated by using the collected data set $\mathcal D_5$ from the previous example as the initial data. In this way, a \emph{knowledge base} is created for a specific material under consideration,  which can be used for further simulations and, at the same time, can be continuously expanded.

The multiscale loop now terminates after only 2 iterations which is due to the described initiation with the available data set. As shown exemplarily for the sectional planes $\bar I_1$-$\bar I_3$ and $\bar I_1$-$\bar I_5$ in Fig.~\ref{fig:12}, a wide range of relevant states is already covered by the states extracted from the cuboid simulation. Thus, the advantage of a transformation into the invariant space becomes again very clear. Although the deformation of the cuboid tensile specimen and torsional specimen is very different in the $\bte C$-space, both overlap clearly in invariant space which is in accordance to \cite{Kalina2021}.
Caused from the geometry and the anisotropic nonlinear elastic behavior, a complex deformation of the macroscopic sample occurs, cf. the surface plot of $\bar C_{12}$ on the deformed geometry in Fig.~\ref{fig:13}.

\subsubsection{Cook's membrane}

\begin{figure}[b!]
	\centering
	\includegraphics{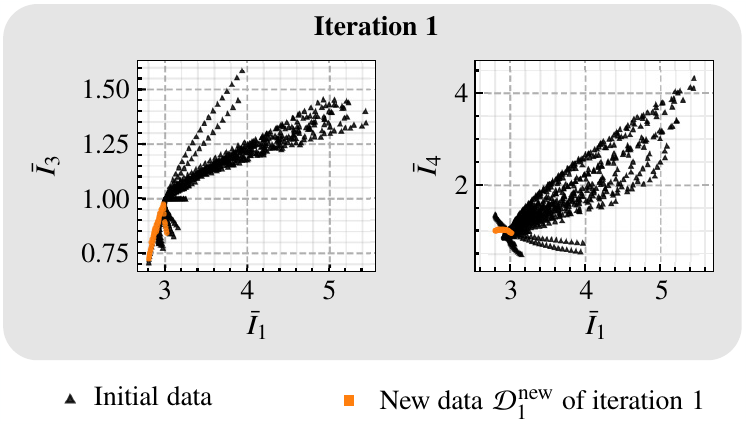}
	\caption{Sampling of the invariant space $\mathcal I^\shortparallel$ within the simulation of the Cook's membrane. The initial data set contains the load cases given in Tab.~\ref{tab:2} as well as the data collected from the cuboid and the torsional sample simulation. The multiscale scheme was terminated in iteration 2. Exemplarily, the sectional planes $\bar I_1$-$\bar I_3$ and $\bar I_1$-$\bar I_4$ are depicted.}
	\label{fig:14}
\end{figure}

\begin{figure*}
	\centering
	\includegraphics{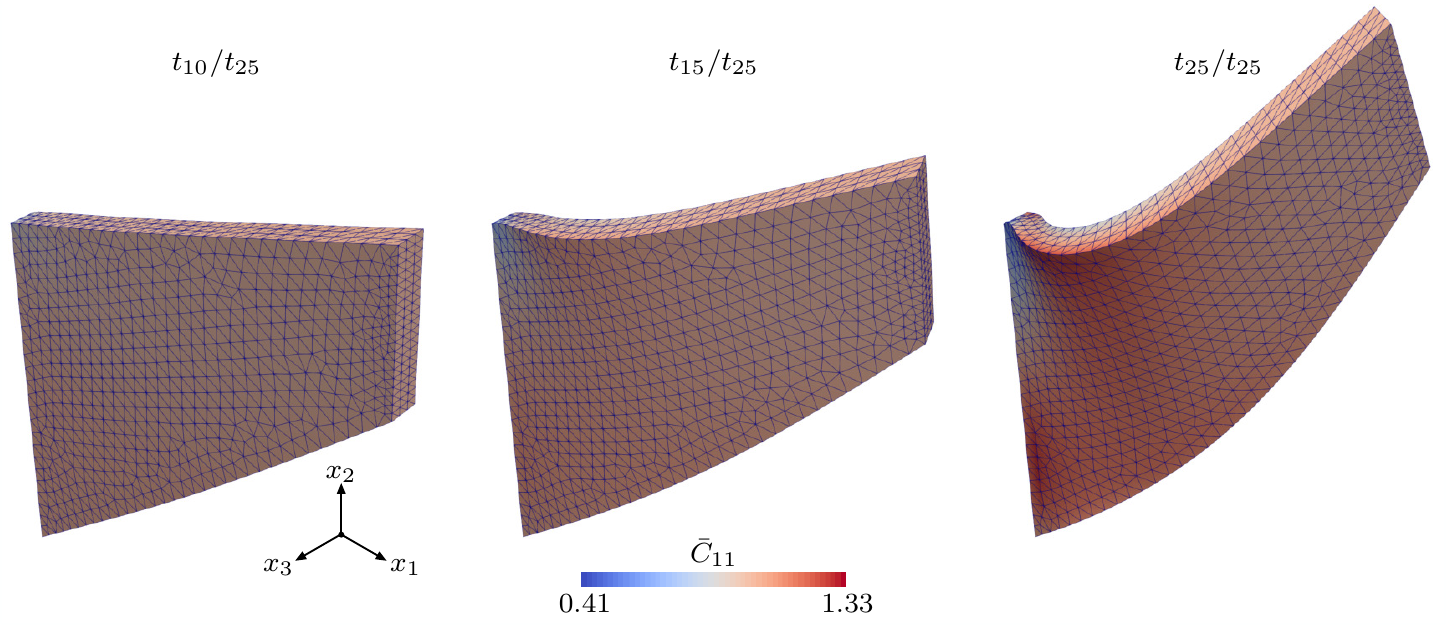}
	\caption{Cook's membrane: distribution of the macroscopic right Cauchy-Green deformation $\bar C_{11}$ at the time steps $t_{10}$, $t_{15}$ and $t_{25}$, where $t_{25}$ is the final time step. The surface plots are given on the deformed configuration $\bar{\mathcal B}$.}
	\label{fig:15}
\end{figure*}	

Finally, as a last example, \emph{Cook's membrane} is simulated by using the developed data-driven multiscale scheme. Thereby, the fiber direction within the membrane is prescribed to $\ve A^\text{macro}=(\ve e_1 + \ve e_3) / \sqrt{2}$. To initiate the scheme, the data set from the previous example is used, where this set also contains data collected within the first example. 

Again, a fast convergence of the multiscale loop is achieved after only 2 overall iterations. 
As shown exemplarily for the sectional planes $\bar I_1$-$\bar I_3$ and $\bar I_1$-$\bar I_5$ in Fig.~\ref{fig:14}, a wide range of relevant states is already covered by the states extracted from the cuboid and the torsional sample simulations. Thus, as already mentioned, the advantage of a transformation into the invariant space is underpinned. Although the deformation of the cuboid tensile specimen, the torsional specimen and the Cook membrane are very different in the $\bte C$-space, they overlap clearly in invariant space.
The deformed macroscopic states with the right Cauchy-Green deformation $\bar C_{11}$ are depicted in Fig.~\ref{fig:15} for the time steps $t_{10}$, $t_{15}$ and $t_{25}$, where the load $\hat{\bve p}$ is applied linear within the steps $t_i\in\{t_0,t_1,\dots,t_{25}\}$. Due to the obliquely oriented fibers with respect to the alignment of the Cook membrane in the $x_1$-$x_2$-plane, an \emph{out of plane deformation} of the Cook membrane occurs. This effect results from the coupling between shear and tension which is a well known effect of fiber reinforced materials.
Although a relatively complex response behavior occurs here, the multiscale problem can be solved quickly and without further human supervision even in this example. 


\section{Conclusions}
\label{sec:Conc}

In this work, a novel \emph{data-driven multiscale approach called FE}${}^\textit{ANN}$ is presented. It is based on \emph{physics-constrained ANNs} which are used as highly efficient surrogate models and an \emph{unsupervised data mining} process. 
The approach allows the efficient simulation of materials with complex underlying microstructures which reveal an overall \emph{anisotropic} and \emph{nonlinear elastic} behavior on the macroscale, e.\,g., composites, architectured materials with pronounced microstructure or foams. The framework has been implemented in such a way, that it is usable on a \emph{HPC cluster} based on the Batch-System \emph{SLURM}. 

Starting from basic kinematics and stress measure definitions, a short revision of anisotropic hyperelastic constitutive models at finite strains is given. Furthermore, a \emph{Hill-type homogenization} framework is described in brief. 
Based on this theoretical basis, the developed \emph{data-driven multiscale scheme} is illustrated in detail. This includes the general procedure and a description of the single steps: initial data generation (a), training process of the ANN (b), macroscopic simulation (c), data analysis (d) and data enrichment (e). Afterwards, the approach is exemplarily applied to the solution of three demonstrative examples, a cuboid under tension, a torsional sample and the Cook membrane. Thereby, the considered macroscopic bodies consist of a \emph{fiber reinforced composite} revealing a highly nonlinear behavior of the individual phases. Due to the incorporation of physical knowledge into the ANN-based surrogate model, only a small number of computationally expensive RVE simulations was needed to solve the considered macroscopic problems. Furthermore, a rather high accuracy of the surrogate model has been shown within a validation.

Altogether, the presented data-driven approach has shown to be an efficient tool for the solution of complex multiscale problems at finite strains. Due to the implemented unsupervised data mining, it is universally applicable to various macroscopic geometries and BCs. 
In order to extend the scheme's application area, several extensions have to be made in the future. For instance, an extension to further \emph{material symmetry groups} \cite{Ebbing2010} have to be made by integrating appropriate invariant sets into the implementation. Finally, an extension to \emph{dissipative constitutive behavior} \cite{Masi2021,Masi2021a,Vlassis2021b} is needed.

\textbf{Acknowledgements: }
All presented computations were performed on a PC-Cluster at the Center for Information Services and High Performance Computing (ZIH) at TU Dresden.
The authors thus thank the ZIH for generous allocations of computer
time. Finally, the authors want to thank Vincent Scholz for several discussions on the topic.

\bibliographystyle{spphys}
{\small \bibliography{bibliography.bib}}

\appendix
\section{Fulfillment of the growth condition by adjusted physics-constrained ANNs}
\label{sec:App2}

Within this appended section, it is discussed how the \emph{growth condition}, i.\,e., 
\begin{align}
	\bar \psi^\text{ANN}(\bte C)\to\infty \; \text{as} \; (\bar J\to\infty \vee \bar J \to 0^+) \; ,
	\label{eq:growth}
\end{align}
can be fulfilled by construction of physics-constrained ANNs. Furthermore, a comparison of such an adapted network with a network neglecting the growth condition is given. 

\subsection{Network architecture and additional constraints}
As discussed in Subsect.~\ref{SubSec:ANN-model}, there are several requirements for the ANN to fulfill Eq.~\eqref{eq:growth} by construction:
First of all, a non-bounded activation function is necessary in any case. This requirement is fulfilled by using the \emph{Softplus} activation function, cf. Eq.~\eqref{eq:softplus}. Secondly, the additional invariant $\bar I_3^* := 1/\bar I_3$ has to be included in the argument list of $\bar \psi^\text{ANN}$. This is to enable the fulfillment of Eq.~\eqref{eq:growth} for $\bar J\to 0^+$. 
Thirdly, additional \emph{constraints} have to be satisfied by the weights belonging to $\bar I_3$ and $\bar I_3^*$.
Accordingly, a possible constraint is given by
\begin{align}
	\begin{split}
		&\left(W_\alpha > 0 \forall \alpha \in \mathcal N\right) \cdots \\ 
		\cdots \wedge &\left(\exists \,w_{\alpha 3}>0 \text{ with } \alpha \in \mathcal N\right) \wedge \left(\exists \,w^*_{\alpha 3}>0 \text{ with } \alpha \in \mathcal N\right) \; ,
		\label{eq:appGrowth}
	\end{split}
\end{align}
which is a sufficient condition to guarantee that Eq.~\eqref{eq:growth} holds. In the equation above, the set $\mathcal N:=\{1,2,,\ldots,N\}$ contains the indices of the hidden layer neurons. 
A proof of Eq.~\eqref{eq:appGrowth} is given in the following. Thereby, taking into account the requirements given above, the following network with one hidden layer is considered:
\begin{align}
	\bar \psi^\text{ANN} := B + \sum_{\alpha=1}^{N} W_{\alpha}  \Softplus\Big(\sum_{\beta=1}^{n} w_{\alpha\beta} \bar{\mathfrak i}_\beta  + \sum_{\beta \in \mathcal A} w^*_{\alpha\beta} \bar{\mathfrak i}^*_\beta +  b_\alpha\Big) \; .
	\label{eq:ANNconvex}
\end{align}

\begin{figure*}
	\centering
	\includegraphics{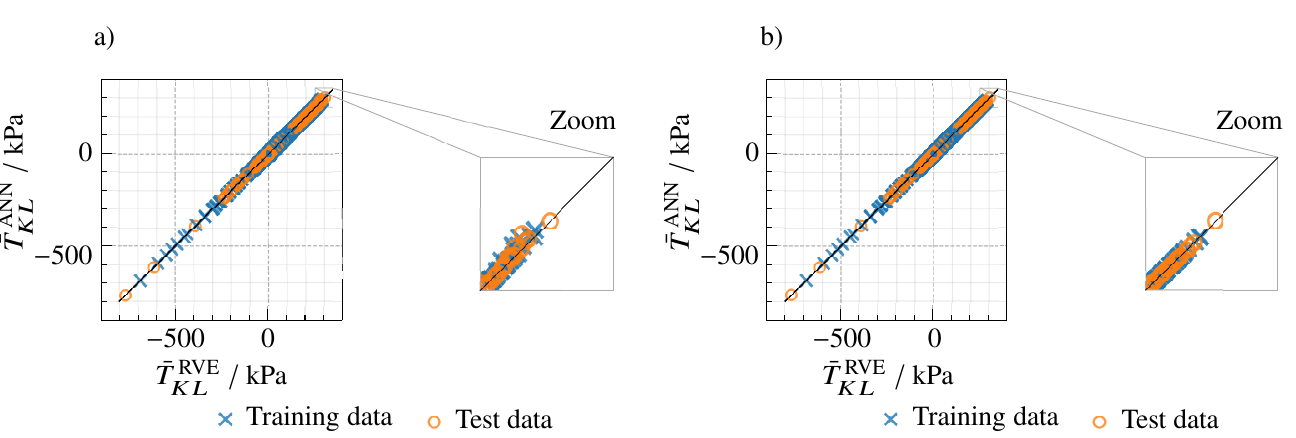}
	\caption{Comparison of an adapted network which fulfills the growth condition \eqref{eq:appGrowth} and a network which neglects this condition: stress predictions of the ANNs and reference stresses obtained from RVE simulations for (a) adapted network and (b) non-adapted network.}
	\label{fig:app_growth}
\end{figure*}

Now, the purely volumetric deformation state $\bte F= \bar \lambda \te 1$ is analyzed to study the model's behavior for $\bar J\to 0^+$ and $\bar J \to \infty$. 
For this state, the relevant invariants follow to
\begin{align}
	\bar I_1 = 3 \bar \lambda^2 \; , \; \bar I_2 = 3 \bar \lambda^4 \; , \; \bar I_3 = \bar \lambda^6 \; , \; \bar I_4=\bar \lambda^2 \; , \; \bar I_5 = \bar \lambda^4 \; , \; \bar I_3^* = \bar \lambda^{-6} \; .
\end{align}
Additionally, the normalization of the invariants according to 
\begin{align}
	\bar{\mathfrak i}_\alpha(\bar I_\alpha) :=\left[
	\bar I_\alpha - \frac{\bar I_\alpha^\text{max}+\bar I_\alpha^\text{min}}{2}\right]
	\frac{2}{\bar I_\alpha^\text{max}-\bar I_\alpha^\text{min}}
	\label{eq:normalization}
\end{align}
has to be taken into account, where $\bar I_\alpha^\text{max} \in \R$ and $\bar I_\alpha^\text{min} \in \R$ denote maximum and minimum components of a given training data set, i.\,e., these values are finite. As one can see from Eq.~\eqref{eq:normalization}, the applied normalization has no influence on the respective power order. 
An evaluation of the invariants for $\bar \lambda \to 0^+$ and $\bar \lambda \to \infty$ gives 
\begin{align}
	\lim_{\bar \lambda \to 0^+} \bar{\mathfrak i}_\alpha = -d_\alpha \; \text{,} \; \lim_{\bar \lambda \to 0^+} \bar{\mathfrak i}^*_3 = \infty \; \text{and} \;
	\lim_{\bar \lambda \to \infty} \bar{\mathfrak i}_\alpha = \infty \; \text{,} \; \lim_{\bar \lambda \to \infty} \bar{\mathfrak i}^*_3 = -d_3^* \; , 
	\label{eq:invariantsLim}
\end{align}
with $d_\alpha,d_3^*\in\R_+$
respectively.  Starting from Eq.~\eqref{eq:ANNconvex} to analyze the case $\bar \lambda \to 0^+$, it follows
\begin{align}
	\begin{split}
		\lim_{\bar \lambda \to 0^+} \bar \psi^\text{ANN} = 
		\lim_{\bar \lambda \to 0^+} \Bigg[ B +  \sum_{\alpha=1}^{N} W_{\alpha}  \log\Big(1+\exp\Big( &\sum_{\beta=1}^{n} w_{\alpha\beta} \bar{\mathfrak i}_\beta + \cdots \\
		& \cdots w^*_{\alpha 3} \bar{\mathfrak i}^*_3 + b_\alpha\Big)\Big) \Bigg] \; .
	\end{split}
\end{align}
Using Eq.~\eqref{eq:invariantsLim} and supposing that it exists at least one $w_{\alpha 3}^*>0$, one finds that
\begin{align}
	\lim_{\bar \lambda \to 0^+} \bar \psi^\text{ANN} = \lim_{\bar \lambda \to 0^+} \bar \lambda^{-6} \frac{2}{\bar I_3^\text{*,max}-\bar I_3^\text{*,min}} \underbrace{\sum_{\alpha=1}^N W_\alpha w^*_{\alpha 3} \theta(w^*_{\alpha 3})}_{C^*_3} \; ,  
\end{align}
where $\theta: \R \to \{0,1\}$ denotes the Heaviside step function. 
Thereby, it has been utilized that the values $d_\alpha$ and $B$, $b_\alpha$ are finite and are negligible with respect to $\bar{\mathfrak i}_3^*$. Consequently, it holds
\begin{align}
	\lim_{\bar \lambda \to 0^+} \bar \psi^\text{ANN} = \infty \; \text{if} \; C^*_3 > 0 \; .
	\label{eq:suff1}
\end{align}
In the same way, supposing that it exists at least one $w_{\alpha 3}>0$, it holds
\begin{align}
	\lim_{\bar \lambda \to \infty} \bar \psi^\text{ANN} 
	= \lim_{\bar \lambda \to \infty} \bar \lambda^{6} \frac{2}{\bar I_3^\text{max}-\bar I_3^\text{min}} \underbrace{\sum_{\alpha=1}^N W_\alpha w_{\alpha 3} \theta(w_{\alpha 3})}_{C_3} \; . 
\end{align}
Consequently, similar to Eq.~\eqref{eq:suff1}, it holds
\begin{align}
	\lim_{\bar \lambda \to \infty} \bar \psi^\text{ANN} = \infty \; \text{if} \; C_3 > 0 \; .
	\label{eq:suff2}
\end{align}
Note that for the case $\bar \lambda \to \infty$, the other invariants which also tend towards infinity have no influence, since $\bar I_3$ is the leading term. 
The above two conditions given in Eqs.~\eqref{eq:suff1} and \eqref{eq:suff2} together constitute a sufficient condition for the growth condition to be satisfied.
However, an easy to implement condition which is also sufficient but more restrictive is given by Eq.~\eqref{eq:appGrowth}. As one can see, it is included in the conditions $C^*_3 > 0 \wedge C_3 > 0$.

\subsection{Comparison of an adapted network and a network neglecting the growth condition}

Here, the network which is used in the examples discussed within Sect.~\ref{sec:4} is compared to an adapted network which, in contrast to the other network, fulfills the growth condition.
To this end, the prediction quality for the data set containing relevant deformation states of the cuboid, the torsional sample and the Cook's membrane is analyzed for both ANNs. Thereby, the constraint \eqref{eq:appGrowth} has been taken into account within the training of the second network.

\begin{table}
	\centering
	\caption{Relative errors of the single components within the stress prediction. ANN-1 and ANN-2 designate the network without and with further constraint \eqref{eq:appGrowth}, respectively.}
	\begin{tabular}{c c c}
		& ANN-1 & ANN-2 \\
		\hline
		$\max(\Delta \bar T_{11}/\bar T_{11}^\text{RVE})$ & $\SI{0.44}{\percent}$ & $\SI{1.29}{\percent}$\\
		$\max(\Delta \bar T_{22}/\bar T_{22}^\text{RVE})$ & $\SI{0.80}{\percent}$ & $\SI{2.75}{\percent}$\\
		$\max(\Delta \bar T_{33}/\bar T_{33}^\text{RVE})$ & $\SI{0.34}{\percent}$ & $\SI{1.58}{\percent}$\\
		$\max(\Delta \bar T_{23}/\bar T_{23}^\text{RVE})$ & $\SI{0.64}{\percent}$ & $\SI{3.38}{\percent}$\\
		$\max(\Delta \bar T_{13}/\bar T_{13}^\text{RVE})$ & $\SI{14.51}{\percent}$ & $\SI{20.62}{\percent}$\\
		$\max(\Delta \bar T_{12}/\bar T_{12}^\text{RVE})$ & $\SI{0.99}{\percent}$ & $\SI{2.31}{\percent}$\\
	\end{tabular}
	\label{tab:app}
\end{table}

The stresses predicted by the networks and the reference stresses obtained from RVE simulations are given in Fig.~\ref{fig:app_growth}(a) and (b). As one can see, the prediction quality is very good for both ANNs. However, regarding the zoom plots, a noticeable difference between both networks becomes apparent. Accordingly, the prediction quality of the adapted network which fulfills the growth condition is declined compared to the network with no further constraints on the weights. This is underlined by a comparison of the maximum relative erros within the stress components given in Tab.~\ref{tab:app}. Note that the comparatively large maximum error in $\bar T_{13}$ results from the small stresses within these component.  

\section{Comparison of hexagonal unit cells and random cells}
\label{sec:App1}

\begin{figure*}
	\centering
	\includegraphics{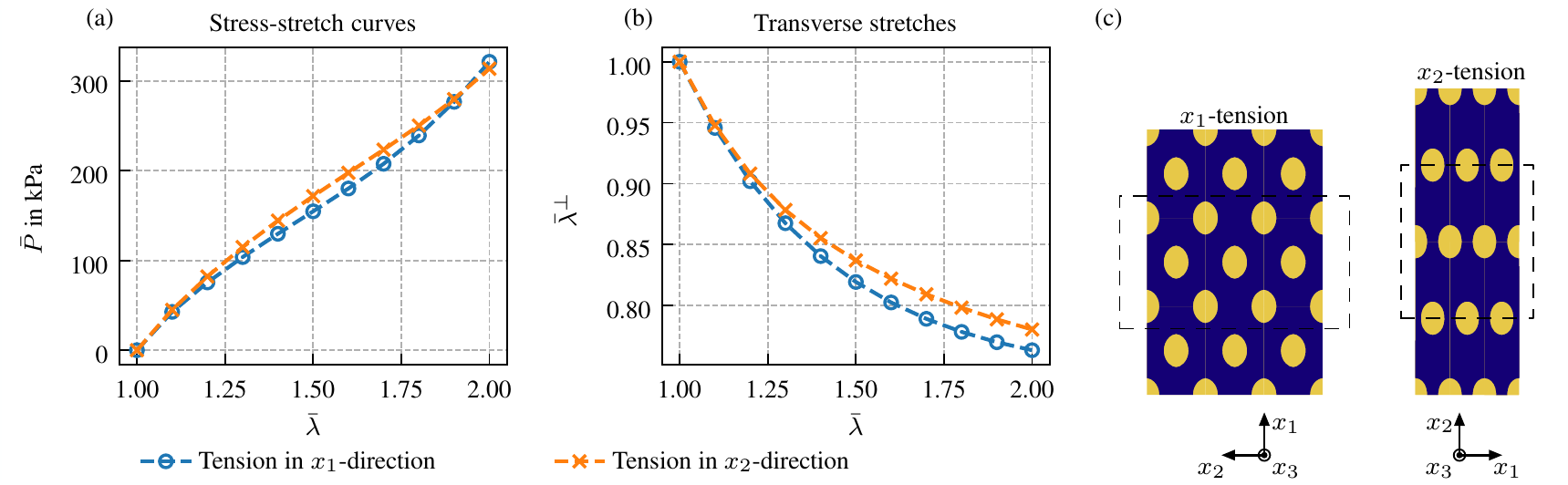}
	\caption{Uniaxial tension of a hexagonal unit cell with fiber orientation in $x_3$-direction: (a) and (b) comparison of stress $\bar P$ and transverse stretch $\bar \lambda^\perp$ for tension in $x_1$- and $x_2$-direction and (c) deformed microstructure combined out of $3\times 3$ and $3\times 2$ periodic unit cells, respectively. The undeformed state is marked with a black dashed line.}
	\label{fig:app1}
\end{figure*}

In this appended section, the effective stress-strain response of a fiber reinforced composite represented by two different microstructures,  an ideal hexagonal and a random distribution, are compared to each other. The first microstructure is thus represented by a unit cell, whereas the second one consists of 100 fibers to capture for statistical effects. The fiber orientation points in the $x_3$-direction for both. 
Exemplarily, a uniaxial tension into the $x_1$- and the $x_2$- direction, i.\,e., perpendicular to the fiber orientation, are considered, where a maximum stretch of $\bar \lambda = 2$ is applied. 

In Fig.~\ref{fig:app1}(a) and (b), the stress-stretch curves and the transverse stretch $\bar \lambda^\perp$ are depicted for the hexagonal unit cell. Thereby, $\bar \lambda^\perp$, which follows due to lateral contraction, is measured in the $x_2$- or $x_1$-direction, respectively. 
Since in the undeformed state all fibers have the same distance in the $x_1$-$x_2$-plane, the curves in the initial region are nearly equivalent. However, as the deformation of the RVE increases, the curves deviate more and more from each other. This is due to the fact that the arrangement of the microstructure changes significantly as a result of the deformation which could be termed a deformation induced anisotropy. 
Regarding the deformed microstructure for tension in the $x_1$-direction, it no longer corresponds to the arrangement in the case of tension in the $x_2$-direction (rotated by 90 degrees), cf. Fig.~\ref{fig:app1}(c). Thus, in summary, the material loses the property of transverse isotropy if finite deformations occur. 

\begin{figure*}
	\centering
	\includegraphics{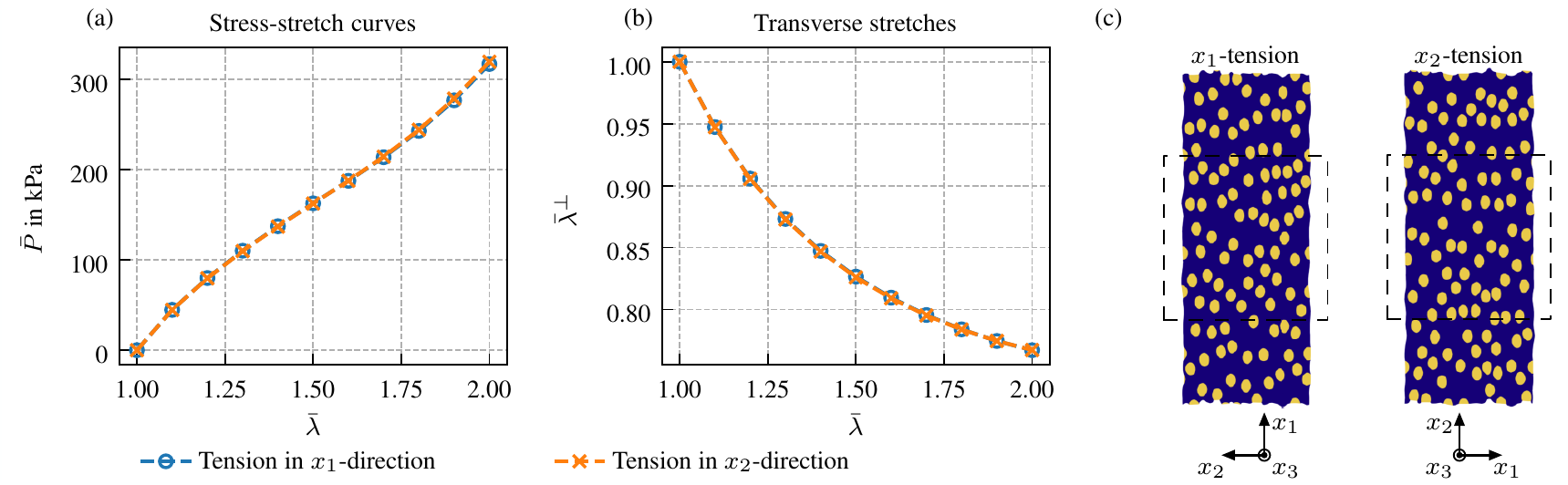}
	\caption{Uniaxial tension of a stochastic RVE with fiber orientation in $x_3$-direction: (a) and (b) comparison of stress $\bar P$ and transverse stretch $\bar \lambda^\perp$ for tension in $x_1$- and $x_2$-direction and (c) deformed microstructure. The undeformed state is marked with a black dashed line.}
	\label{fig:app2}
\end{figure*}

Compared to this, the same uniaxial loadings are depicted in Fig.~\ref{fig:app2}(a) and (b) for the cell with a random fiber distribution. As one can see there, the curves lie on top of each other over the complete range of stretch $\bar \lambda \in[1,2]$. Thus, in contrast to the hexagonal unit cell, a transversely isotropic effective behavior -- which is expected for a fiber reinforced composite -- results even for finite strains, whereby the \mbox{$x_1$-$x_2$}-plane is the isotropy plane. In order to simulate the overall behavior of a realistic fiber reinforced composite, the usage of a statistical RVE is thus mandatory for finite strains.

\end{document}